\documentclass[journal]{IEEEtran}

\usepackage{cite}
\usepackage{amsmath,amssymb,amsfonts}
\usepackage{graphicx}
\usepackage{textcomp}
\usepackage{xcolor}
\usepackage{subfigure}
\usepackage{bm}
\usepackage{bbm}

\usepackage{booktabs}
\usepackage{algorithm}
\usepackage{amsmath}
\usepackage[noend]{algpseudocode}
\usepackage{multirow}

\begin{document}

\title{Quaternion Factorization Machines: \\A Lightweight Solution to Intricate Feature Interaction Modelling}

\author
{
 Tong~Chen, Hongzhi~Yin, Xiangliang~Zhang, Zi~Huang, Yang~Wang, Meng~Wang\\
  \IEEEcompsocitemizethanks{
  \IEEEcompsocthanksitem T. Chen, H. Yin, and Z. Huang are with the School of Information Technology and Electrical Engineering, The University of Queensland.\protect\\
E-mail: tong.chen@uq.edu.au, h.yin1@uq.edu.au, huang@itee.uq.edu.au
\IEEEcompsocthanksitem X. Zhang is with the Machine Intelligence and Knowledge Engineering Laboratory, King Abdullah University of Science and Technology.\protect\\
E-mail: xiangliang.zhang@kaust.edu.sa
\IEEEcompsocthanksitem Y. Wang and M. Wang are with the Key Laboratory of Knowledge Engineering with Big Data, Ministry of Education, School of Computer Science and Information Engineering, Hefei University of Technology, China.\protect\\
E-mail: yangwang@hfut.edu.cn, eric.mengwang@gmail.com

}
\thanks{Manuscript is under review. Hongzhi Yin is the corresponding author.}
}


\IEEEtitleabstractindextext{%
\begin{abstract}
Due to the sparsity of available features in web-scale predictive analytics, combinatorial features become a crucial means for deriving accurate predictions. As a well-established approach, factorization machine (FM) is capable of automatically learning high-order interactions among features to make predictions without the need for manual feature engineering. With the prominent development of deep neural networks (DNNs), there is a recent and ongoing trend of enhancing the expressiveness of FM-based models with DNNs. However, though better results are obtained with DNN-based FM variants, such performance gain is paid off by an enormous amount (usually millions) of excessive model parameters on top of the plain FM. Consequently, the heavy parameterization impedes the real-life practicality of those deep models, especially efficient deployment on resource-constrained IoT and edge devices.

In this paper, we move beyond the traditional real space where most deep FM-based models are defined, and seek solutions from quaternion representations within the hypercomplex space. Specifically, we propose the quaternion factorization machine (QFM) and quaternion neural factorization machine (QNFM), which are two novel lightweight and memory-efficient quaternion-valued models for sparse predictive analytics. By introducing a brand new take on FM-based models with the notion of quaternion algebra, our models not only enable expressive inter-component feature interactions, but also significantly reduce the parameter size due to lower degrees of freedom in the hypercomplex Hamilton product compared with real-valued matrix multiplication. Extensive experimental results on three large-scale datasets demonstrate that QFM achieves $4.36\%$ performance improvement over the plain FM without introducing any extra parameters, while QNFM outperforms all baselines with up to two magnitudes' parameter size reduction in comparison to state-of-the-art peer methods.
\end{abstract}

}

\maketitle

\IEEEdisplaynontitleabstractindextext

%
\IEEEpeerreviewmaketitle


\section{Introduction}\label{sec:intro}

\IEEEPARstart{I}{n} the context of big data, predictive analytics play a pivotal role in various applications, where the prediction targets range from user ratings \cite{wu2017recurrent,chen2020sequence} to product sales \cite{chen2018tada} and traffic conditions \cite{wang2019origin}. Essentially, the goal of predictive analytics is to learn a function that accurately maps the input variables (i.e., features) to the desired output. With its straightforward connection to business revenue and user experience, immense research attention has been paid to improving the quality of web-scale predictions. 

Witnessed by various prior efforts in this area, modelling the interactions among different features acts as a winning formula for a wide range of prediction tasks \cite{lian2017practical,chen2020try,shan2016deep}. The interactions among multiple raw features are usually termed as \textbf{\textit{cross features}} \cite{shan2016deep} (a.k.a. multi-way features and combinatorial features). On one hand, in order to deal with the ubiquitous categorical features (e.g., user IDs), a common practice is to covert them into one-hot encodings \cite{he2017neuralcol,shan2016deep,rendle2011fast,chen2019air} to make traditional prediction methods like logistic regression \cite{hosmer2013applied} and support vector machine \cite{chang2011libsvm} applicable. Since these one-hot features are usually of high dimensionality but sparse owing to the large number of possible category variables \cite{he2017neural}, directly using raw features leads to severe overfitting \cite{song2019autoint} and rarely provides optimal results. On the other hand, the combinatorial effect of cross features creates a pathway to augmenting the contextual information provided by raw features. For example, when predicting a user's income, coupling her/his occupation with age can offer indicative signals of this user's seniority on a job position, e.g., $[occupation\, =\, engineer,\, age \,=\, 38]$, thus yielding richer semantics and better prediction accuracy. 

For different data mining tasks, handcrafting effective cross features requires extensive domain knowledge and is labor-intensive, and the designed cross features can hardly generalize to new tasks and domains. To avoid the high cost of task-specific feature engineering, \textbf{\textit{factorization machine}} (FM) \cite{rendle2010factorization} is proposed to embed raw features into a latent space, and model the second-order interactions among features via the inner product of their embedding vectors. To better capture the effect of feature interactions, variants of the plain FM are proposed, like FFM \cite{juan2016field} that models feature interactions in field-aware embedding spaces and HOFM \cite{blondel2016higher} for modelling higher-order feature interactions. 

Despite the appealing simplicity and compatibility of these linear FM-based approaches, their performance is largely constrained by the limited expressiveness \cite{he2017neural} when modelling the subtle and complex feature interactions. Recently, motivated by the capability of learning discriminative representations from raw inputs, deep neural networks (DNNs) \cite{lecun2015deep} have been adopted to enhance the modelling of feature interactions. For instance, He \textit{et al.} \cite{he2017neural} bridges the cross feature scheme of FM with the non-linearity of DNN, and proposes a neural factorization machine (NFM). Instead of the straightforward inner product in FM, NFM takes the sum of all features' linear pairwise combinations into a feed-forward neural network, and generates a latent representation of high-order feature interactions. With the idea of learning high-order feature interactions with DNNs, various DNN-based approaches are devised for predictive analytics \cite{xiao2017attentional,cheng2016wide,lian2018xdeepfm,shan2016deep,guo2017deepfm,qu2016product,chen2019air} and demonstrate advantageous prediction performance. 

However, compared with the plain FM, all the aforementioned DNN-based variants are heavily parameterized and thus memory-inefficient. Consequently, the achieved performance gain comes at a high cost of introducing excessive parameters in their neural architectures. For instance, the optimal performance of xDeepFM \cite{lian2018xdeepfm} is commonly achieved with a configuration of at least three convolution layers joined by two feed-forward deep layers \cite{liu2019feature,song2019autoint,lian2018xdeepfm}, leading to millions of additional model parameters on top of the plain FM (see Section \ref{sec:exp}). As a result, the bulkiness of existing deep FM-based models inevitably renders practical and efficient model deployment challenging. These memory-intensive models are also too cumbersome to welcome the emerging opportunities of applications on mobile or IoT devices \cite{wang2020next}. Furthermore, as another side effect, the excessive parameterization creates an obstacle for effectively learning useful interaction patterns from a large number of parameters, especially with sparse features \cite{liu2019feature}. As more sophisticated DNN structures are being utilized to model feature interactions (e.g., the graph neural networks in \cite{li2019fi} and transformers in \cite{song2019autoint}), enhancing FM with DNNs appears to be an ongoing trend. In light of this, it is crucial to derive a new paradigm that only requires lightweight parameterization just like the plain FM, while allowing for comprehensive feature interaction modelling to generate accurate predictions. 

Before taking the step to derive a corresponding solution, a key observation we have drawn in this line of research is that most work has primarily focused on real-valued representations in the $\mathbb{R}$ space. In contrast, building machine learning models with the notion of quaternion algebra \cite{kuipers1999quaternions} in the hypercomplex space $\mathbb{H}$ has just started revealing its immense potential through recent advances \cite{parcollet2019survey}. In quaternion algebra, the Hamilton product plays a key role in supporting intricate interactions between quaternion representations while minimizing the parameter consumption. Compared with the matrix multiplication that is common in all DNN-based models' deep layers \cite{he2017neural,shan2016deep,lian2018xdeepfm,liu2019feature}, the Hamilton product between two quaternions has fewer degrees of freedom, enabling up to four times compression of parameter size in a single multiplication operation. Moreover, quaternions are hypercomplex numbers consisting of one real and three imaginary components, in which the inter-dependencies between these components are naturally encoded during training via the Hamilton product \cite{tay2019lightweight}. Thus, machine learning approaches that utilize quaternion-valued representations and operations are inherently advantageous in modelling the subtle yet complex interactions between features, reflected by an emerging surge in successful applications like image processing \cite{gaudet2018deep,parcollet2019quaternion_CNN}, speech recognition \cite{parcollet2019quaternion_RNN}, and text classification \cite{tay2019lightweight}. 
More recently, \cite{zhang2019quaternion_CF} successfully rolls out a quaternion collaborative filtering framework by exploiting the computation of quaternion algebra to quantify the affinity between users and items. Hence, this verifies the capability of quaternions in modelling subtle interactions among latent representations, which is also the essence of the plain FM and its successors. 

To this end, to bypass the limitations of using real-valued DNNs for feature interaction modelling, we move beyond the real space, and subsume FM-based models under a quaternion-valued learning framework for the first time. In this paper, we propose the first \textit{\textbf{quaternion factorization machine}} (QFM) that generalizes the original FM into the hypercomplex domain, and then further infuse non-linear expressiveness into QFM by extending it to a deep neural model, namely \textit{\textbf{quaternion neural factorization machine}} (QNFM). By making full use of quaternions, we create an ideal bootstrap between a compact model size and uncompromized performance. In both models, we parameterize the feature interaction schemes as quaternion-valued functions in the hypercomplex space $\mathbb{H}$, and each feature embedding diverges into four quaternion components. In QFM, to thoroughly model feature interactions, we replace the shallow dot product in the plain FM with our proposed inner Hamilton product for quaternion vector pairs. On this basis, we further propose an innovative asymmetrical quaternion interaction pooling operation in QNFM to encode the latent information of all pairwise feature interactions, followed by a quaternion-valued feed-forward network that learns the complex and high-order interaction signals. Intuitively, the utilization of Hamilton product in QFM and QNFM is in similar spirit to multi-view representations \cite{xu2013survey}, where the interactions between features are evoked by the connectivity and information exchange among the four quaternion components in the hypercomplex space. As a result, this leads to an expressive blend of contexts when forming the final representations, yielding accurate predictions without the need for overwhelmingly sophisticated neural networks and excessive model parameters. 

This paper brings the following contributions:
\begin{itemize}
	\item We address a common drawback of existing DNN-based FM variants, i.e., the heavy parameterization inherited from real-valued neural networks. We sought solutions from the hypercomplex space by exploiting quaternion representations. To the best of our knowledge, this is the first exploration of quaternion FM-based models.
	\item We propose QFM and QNFM, two novel and lightweight methods for predictive analytics. By utilizing quaternion representations and defining interaction paradigms with the notion of Hamilton product, our models are highly expressive when modelling the intricate feature interactions, and can greatly reduce the parameter consumption.
	\item We extensively evaluate our models on three large-scale real datasets. Experiments show that QFM yields considerable improvements over the plain FM with no extra parameter costs, while QNFM achieves the state-of-the-art prediction accuracy with a parameter reduction by up to two magnitudes regarding its real-valued counterparts.
\end{itemize}

\section{Preliminaries}
\subsection{Notations} 
Throughout this paper, all vectors and matrices are respectively denoted by bold lower case and bold upper case letters, e.g., $\mathbf{x}$ and $\mathbf{X}$, and all vectors are \textit{\textbf{column vectors}} unless specified. To maintain the simplicity, we use superscript $\diamond$ to distinguish all quaternion-valued parameters. Sets are represented by calligraphic uppercase letters, e.g., $\mathcal{V}$. Hollowed uppercase letters $\mathbb{H}$ and $\mathbb{R}$ are respectively sets of quaternions and real numbers, while we use $\mathbb{I}$, $\mathbbm{J}$, $\mathbbm{K}$ to denote three imaginary numbers composing a quaternion.

\subsection{Factorization Machines}
Factorization machines (FMs) \cite{rendle2010factorization} are originally proposed for collaborative recommendation. Take instance $[user \, ID\!=\!2, gender\!=\!female, movie \!= \!Avengers, movie \,\, type \!=\! action]$ as an example, its input is a high-dimensional sparse feature $\mathbf{x} \in \{0,1\}^{n}$ constructed by concatenating the one-hot encodings from all feature fields \cite{lian2018xdeepfm,shan2016deep,qu2016product}:
\begin{equation}\label{eq:input}
	\mathbf{x} = \underbrace{[0,1,0,...,0]}_\textnormal{\small user ID} \underbrace{[1,0]}_\textnormal{\small gender} \underbrace{[0,1,0,...,0]}_\textnormal{\small movie} \underbrace{[0,1,0,...,0]}_\textnormal{\small movie type},
\end{equation}
where any real-valued feature (e.g., age) can also be directly included in $\mathbf{x}$ \cite{he2017neural,rendle2011fast}. Then, FMs are linear predictors that estimate the desired output by modelling all interactions between each pair of features within $\mathbf{x}$ \cite{rendle2010factorization}:
\begin{equation}\label{eq:FM}
	\widehat{y} = w_0 + \sum^n_{i=1}{w_ix_i} + \sum^n_{i=1}\sum^n_{j=i+1}{\mathbf{v}_i^{\top} \mathbf{v}_j \cdot x_ix_j},
\end{equation}
where $n$ is the total number of features, $w_0$ is the global bias, $w_i$ is the weight assigned to the $i$-th feature, and $x_i$, $x_j\in \mathbf{x}$ are respectively the observed values of the $i$-th and $j$-th features. $\mathbf{v}_i$, $\mathbf{v}_j \!\in\! \mathbb{R}^{d}$ are corresponding embedding vectors for features $i$ and $j$, while $d$ is the embedding dimension. Thus, the first two terms in Eq.(\ref{eq:FM}) can be viewed as a linear regression scheme, while the third term models the effect of pairwise feature interactions \cite{xiao2017attentional}.

\subsection{Quaternion Algebra}
Quaternion is an extension of complex number that operates on a $4$-dimensional hypercomplex space\footnote{In this paper, we use the terms ``hypercomplex space'' and ``quaternion space'' interchangeably.}. In the Hamilton quaternion space $\mathbb{H}$, a quaternion $q^{\diamond}$ consists of one real part and three imaginary parts:
\begin{equation}
	q^{\diamond} = r1 + a\mathbb{I} + b\mathbb{J} + c\mathbb{K},
\end{equation}
where $r$, $a$, $b$, $c \in \mathbb{R}$ are real numbers, while $1$, $\mathbb{I}$, $\mathbb{J}$ and $\mathbb{K}$ are the quaternion unit basis. In the rest of our paper, we term the four components $r$, $a$, $b$, $c$ as \textbf{\textit{quaternion cores}} and simplify the real part $r1$ as $r$ to be concise. $\mathbb{I}$, $\mathbb{J}$ and $\mathbb{K}$ are imaginary numbers and satisfy the following Hamilton's rule:
\begin{equation}\label{eq:Hamilton_rule1}
	\mathbb{I}^2 = \mathbb{J}^2 = \mathbb{K}^2 = \mathbb{I}\mathbb{J}\mathbb{K} = -1,
\end{equation}
from which multiple useful properties can be derived, e.g., $\mathbb{I}^{2}\mathbb{J}\mathbb{K} = -\mathbb{I}$ leads to $\mathbb{J}\mathbb{K}=\mathbb{I}$. To list a few:
\begin{equation}\label{eq:Hamilton_rule2}
	\mathbb{I}\mathbb{J}=\mathbb{K}, \mathbb{J}\mathbb{I}=-\mathbb{K}, \mathbb{J}\mathbb{K}=\mathbb{I}, \mathbb{K}\mathbb{J}=-\mathbb{I}, \mathbb{K}\mathbb{I}=\mathbb{J}, \mathbb{I}\mathbb{K}=-\mathbb{J}.
\end{equation}

In $\mathbb{H}$, a normalized quaternion, i.e., unit quaternion $\overline{q}^{\diamond}$ is expressed as:
\begin{equation}\label{eq:norm_1}
	\overline{q}^{\diamond} = \frac{q^{\diamond}}{|q^{\diamond}|} = \frac{q}{\sqrt{r^2+a^2+b^2+c^2}}.
\end{equation}
The multiplication, i.e., Hamilton product of two quaternions $q_1 = r_1 + a_1\mathbb{I} + b_1\mathbb{J} + c_1\mathbb{K}$ and $q_2 = r_2 + a_2\mathbb{I} + b_2\mathbb{J} + c_2\mathbb{K}$ can be computed following the distributive law:
\begin{equation}\label{eq:product_1}
\begin{split}
	q_1^{\diamond} \times q_2^{\diamond} & = (r_1r_2 - a_1a_2 - b_1b_2 - c_1c_2)\\
	& + (r_1a_2 + a_1r_2 + b_1c_2 - c_1b_2)\mathbb{I}\\
	& + (r_1b_2 - a_1c_2 + b_1r_2 + c_1a_2)\mathbb{J}\\
	& + (r_1c_2 + a_1b_2 - b_1a_2 + c_1r_2)\mathbb{K},\\
\end{split}
\end{equation}
where we should note that based on the rules in Eq.(\ref{eq:Hamilton_rule1}) and Eq.(\ref{eq:Hamilton_rule2}), the Hamilton product is associative but non-commutative, i.e., $q_1^{\diamond} \times q_2^{\diamond} \neq q_2^{\diamond} \times q_1^{\diamond}$.

\begin{figure*}[!t]
\center
\includegraphics[width = 6in]{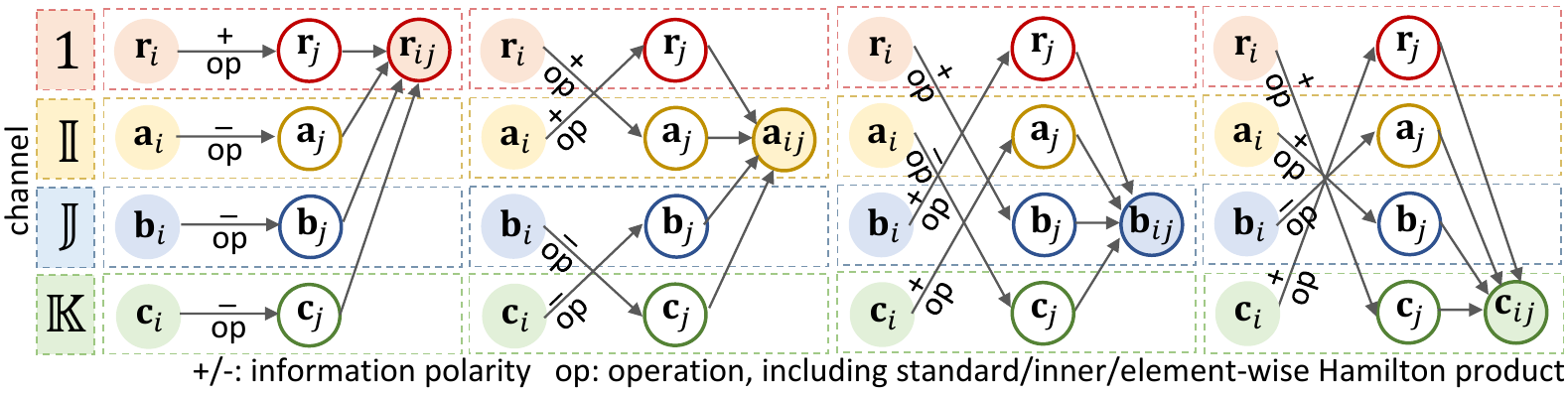}
\vspace{-0.4cm}
\caption{The information flow in a Hamilton product operation. Note that the dimension of four output quaternion cores depends on the exact operation performed, e.g., $r_{ij}, a_{ij}, b_{ij}, c_{ij} \in \mathbb{R}$ in inner Hamilton product while $\mathbf{r}_{ij}, \mathbf{a}_{ij}, \mathbf{b}_{ij}, \mathbf{c}_{ij} \in \mathbb{R}^{d}$ in element-wise Hamilton product.}
\label{Figure:Struct}
\end{figure*}

\section{Quaternion Factorization Machines}
To start with, we firstly introduce the basic form of QFM by partially transferring FM to the quaternion space. In QFM, apart from the real-valued sparse input $\mathbf{x}$ and output $\widehat{y}$, we retain the first two regression terms in Eq.(\ref{eq:FM}) in the real space, while all model parameters and intermediate computations for the feature interactions are defined in the quaternion space. Mathematically, QFM is written as:
\begin{equation}\label{eq:QFM}
	\widehat{y}= w_0 + \sum_{i=1}^{n}w_i x_i + \phi(\mathcal{V}_x^{\diamond}),
\end{equation}
where $w_0$, $w_i \in \mathbb{R}$ are real-valued global bias and weight associated with the $i$-th feature. $\mathcal{V}_x^{\diamond}$ is the set of quaternion embedding vectors for all \textbf{\textit{non-zero features}} in $\mathbf{x}$, $\phi(\cdot)$ is a quaternion-valued function that firstly quantifies the intensity of all pairwise feature interactions in the $\mathbb{H}$ space, and maps the quaternion results into real numbers. In what follows, we introduce the design of $\phi(\cdot)$ in detail.
\subsection{Quaternion Embeddings}
In FM-based methods, the real-valued features are converted into $d$-dimensional vector representations (a.k.a. embedding vectors) to allow for the modelling of pairwise feature interactions. In the quaternion space, we embed features with quaternion vectors. Based on the construction rule of $\mathbf{x} \in \mathbb{R}^n$, it can be viewed as the additive form of the one-hot encodings for all non-zero features. Thus, $\mathbf{x} = \sum _{i=1}^{n}{x_i \mathbf{e}_i}$, where $\mathbf{e}_i = [0,...,0,1,0,...,0]$ is an $n$-dimensional one-hot vector of an individual non-zero feature. Specifically, we define a quaternion embedding vector $\mathbf{v}^{\diamond} \in \mathbb{H}^d$ as $\mathbf{v}^{\diamond} = \mathbf{r} + \mathbf{a}\mathbb{I} +  \mathbf{b}\mathbb{J} + \mathbf{c}\mathbb{K}$, where $\mathbf{r}$, $\mathbf{a}$, $\mathbf{b}$, $\mathbf{c} \in \mathbb{R}^{d}$ are real-valued vectors. Then, our quaternion embedding system consists of four distinct embedding matrices to match up with four quaternion cores in $\mathbf{v}^{\diamond}_i$ for the $i$-th feature:
\begin{align}
\label{eq:embedding}
	\!\!\mathbf{v}_i^{\diamond} &= \mathbf{r}_i + \mathbf{a}_i\mathbb{I} +  \mathbf{b}_i\mathbb{J} + \mathbf{c}_i\mathbb{K}\\
	&= x_i\mathbf{M}_{r}^{\top}\!\mathbf{e}_i + (x_i\mathbf{M}_{a}^{\top}\!\mathbf{e}_i)\mathbb{I} + (x_i\mathbf{M}_{b}^{\top}\!\mathbf{e}_i)\mathbb{J} + (x_i\mathbf{M}_{c}^{\top}\!\mathbf{e}_i)\mathbb{K},\nonumber
\end{align}
where $\mathbf{M}_{r}$, $\mathbf{M}_{a}$, $\mathbf{M}_{b}$, $\mathbf{M}_{c} \in \mathbb{R}^{n\times d}$ are four corresponding embedding matrices, while each embedding component is weighted by the scalar value of feature $x_i$. Due to the sparse nature of $\mathbf{x}$, we only need to include the embeddings of non-zero features in $\mathcal{V}_x^{\diamond}$, i.e., $\mathcal{V}_x^{\diamond} = \{\mathbf{v}^{\diamond}_i | x_i \neq 0 \}_{i=1}^n$ for subsequent computations.  

\subsection{Feature Interaction Modelling with Inner Hamilton Product}
With all quaternion feature embeddings $\mathcal{V}_x^{\diamond}$, we propose a pairwise feature interaction scheme with an \textbf{\textit{inner Hamilton product}} operation that we have defined below to enable cross-features in the quaternion space:
\begin{equation}\label{eq:twoway_interaction}
\begin{split}
		h^{\diamond} & = \sum_{i=1}^{n} \sum_{j=1}^{n} \mathbf{v}^{\diamond}_i \otimes \mathbf{v}^{\diamond}_j\\
		& = \sum_{i=1}^{n} \sum_{j=i+1}^{n} (\mathbf{v}_i^{\diamond} \otimes \mathbf{v}_j^{\diamond} + \mathbf{v}_j^{\diamond} \otimes \mathbf{v}_i^{\diamond}),
\end{split}
\end{equation}
where $h^{\diamond} = r_h + a_h\mathbb{I} + b_h\mathbb{J} + c_h\mathbb{K} \in \mathbb{H}$ is a quaternion that encodes all the pairwise interaction information among features in $\mathcal{V}_x^{\diamond}$. $\otimes$ denotes our proposed inner Hamilton product of two quaternion vectors. For an arbitrary pair of quaternion embeddings $\mathbf{v}^{\diamond}_i$, $\mathbf{v}^{\diamond}_j \in \mathbb{H}^d$, $\mathbf{v}_i^{\diamond} \otimes \mathbf{v}_j^{\diamond}$ is computed as follows:
\begin{equation}\label{eq:product_2}
\begin{split}
	\mathbf{v}_i^{\diamond} \otimes \mathbf{v}_j^{\diamond} & = (\mathbf{r}_i^{\top}\mathbf{r}_j - \mathbf{a}_i^{\top}\mathbf{a}_j - \mathbf{b}_i^{\top}\mathbf{b}_j - \mathbf{c}_i^{\top}\mathbf{c}_j)\\
	& + (\mathbf{r}_i^{\top}\mathbf{a}_j + \mathbf{a}_i^{\top}\mathbf{r}_j + \mathbf{b}_i^{\top}\mathbf{c}_j - \mathbf{c}_i^{\top}\mathbf{b}_j)\mathbb{I}\\
	& + (\mathbf{r}_i^{\top}\mathbf{b}_j - \mathbf{a}_i^{\top}\mathbf{c}_j + \mathbf{b}_i^{\top}\mathbf{r}_j + \mathbf{c}_i^{\top}\mathbf{a}_j)\mathbb{J}\\
	& + (\mathbf{r}_i^{\top}\mathbf{c}_j + \mathbf{a}_i^{\top}\mathbf{b}_j - \mathbf{b}_i^{\top}\mathbf{a}_j + \mathbf{c}_i^{\top}\mathbf{r}_j)\mathbb{K},\\
\end{split}
\end{equation}
where the two quaternion cores interact with each other via inner product, thus yielding a quaternion scalar as the output. It is worth noting that unlike the traditional pairwise feature interaction scheme in real-valued FMs (i.e., $\mathbf{v}_i^{\top}\mathbf{v}_j$ in Eq.(\ref{eq:FM})), we allow the $i$-th and $j$-th features to interact with each other twice, i.e., $\mathbf{v}_i^{\diamond} \otimes \mathbf{v}_j^{\diamond}$ and $\mathbf{v}_j^{\diamond} \otimes \mathbf{v}_i^{\diamond}$. This is because our proposed inner Hamilton product inherits the non-commutative nature from the Hamilton product, making $\mathbf{v}_i^{\diamond} \otimes \mathbf{v}_j^{\diamond}$ and $\mathbf{v}_j^{\diamond} \otimes \mathbf{v}_i^{\diamond}$ non-interchangeable and carry asymmetric interaction contexts. Correspondingly, instead of the straightforward vector inner product, Eq.(\ref{eq:twoway_interaction}) allows for effective infusion of bilateral pairwise feature interactions, thus providing richer predictive power in return. 

\subsection{Rationale of Hamilton Product in Feature Interactions}\label{sec:rationale}
The quaternion representation $\mathbf{v}^{\diamond}_i\in\mathbb{H}^{d}$ of feature $i$ can be thought of as a combination of four information channels, i.e., $1$, $\mathbb{I}$, $\mathbb{J}$, $\mathbb{K}$. Owing to the fact that $\mathbf{v}^{\diamond}_i \in \mathbb{H}^d$ carries four distinct $d$-dimensional real vectors, it can be viewed as evenly dividing a $4d$-dimensional vector $[\mathbf{r}_i;\mathbf{a}_i;\mathbf{b}_i;\mathbf{c}_i]$ into a $4$-channel format. Each information channel carries a real-valued, $d$-dimensional dense vector $\mathbf{r}_i$, $\mathbf{a}_i$, $\mathbf{b}_i$, and $\mathbf{c}_i$. Hence, the inner Hamilton product $\otimes$ between two quaternion embeddings creates a cross-channel interaction scheme following Hamilton's rule. As illustrated in Figure~\ref{Figure:Struct}, when computing $r_{ij} + a_{ij}\mathbb{I} +  b_{ij}\mathbb{J} + c_{ij}\mathbb{K} = \mathbf{v}^{\diamond}_i \otimes \mathbf{v}^{\diamond}_j$, the interaction response in each channel is generated by applying different combinatorial rules to the dot products of real-valued cores in $\mathbf{v}^{\diamond}_i$ and $\mathbf{v}^{\diamond}_j$. Intuitively, the quaternion scalar resulted from inner Hamilton product compresses the signal of cross-channel information flows from one feature to the other, representing their pairwise interaction intensity. 

\subsection{Generating Predictions}
In $\phi(\cdot)$, the last step is to map the quaternion output $h^{\diamond}$ back to the real space, so that the learned feature interaction information can be coupled to the real-valued linear regression term to generate accurate predictions. Since each core in $h^{\diamond}$ can be viewed as an interaction score learned in a distinct channel, we adopt average pooling to merge the information from all channels:
\begin{equation}\label{eq:QFM_pred}
		\phi(\mathcal{V}_x^{\diamond}) = f_{\mathbb{H}\mapsto \mathbb{R}}(h^{\diamond}) = \frac{r_h + a_h + b_h + c_h}{4},
\end{equation}
which will be used in Eq.(\ref{eq:QFM}) to emit the final prediction $\hat{y}$.

\section{Extending QFM: Quaternion Neural \\Factorization Machines}
To harness the advantages of nonlinearity in feature interaction modelling and strengthen the expressiveness of QFM, we further propose quaternion neural factorization machine (QNFM), a neural extension of QFM. Specifically, in QNFM, we preserve the first two linear regression terms in Eq.(\ref{eq:QFM}), and infuse nonlinearity by replacing the linear interaction function $\phi(\cdot)$ with a quaternion neural network-based architecture $\phi'(\cdot)$. The output of QNFM is then $\widehat{y}= w_0 + \sum_{i=1}^{n}w_i x_i + \phi'(\mathcal{V}_x^{\diamond})$. In this section, we present our pathway to building the function $\phi'(\cdot)$ in QNFM.

\subsection{Asymmetrical Quaternion Interaction Pooling}
Similar to FM-based models built upon real-valued deep neural networks, all quaternion feature embeddings in QNFM need to be combined into one unified representation before they are fed into the neural networks to model the complex and non-linear feature interactions. Common approaches include concatenations \cite{lian2018xdeepfm,shan2016deep} and pooling operations \cite{he2017neural,xiao2017attentional} for all feature embeddings. As concatenating all feature embeddings will produce a high-dimensional vector, it will lead to the huge dimensionality of corresponding network parameters (i.e., weights and biases) and increased model sizes. Also, since $\mathcal{V}_x^{\diamond}$ only contains the embeddings of non-zero features, the dimensions of resulted concatenations may vary for different feature embedding sets $\mathcal{V}_x^{\diamond}$ in some cases, making the subsequent neural network ill-posed. On this occasion, pooling operation is a better fit for QNFM as it aims to compress all feature embeddings into a compact and fix-size vector. Thus, we propose \textbf{\textit{asymmetrical quaternion interaction pooling}} in QNFM to merge all embeddings in $\mathcal{V}_x^{\diamond}$ into a condensed quaternion vector representation $\widetilde{\mathbf{v}}^{\diamond} \in \mathbb{H}^{d}$:
\begin{equation}\label{eq:pooling}
\begin{split}
		\widetilde{\mathbf{v}}^{\diamond} 
		&= \sum_{i=1}^{n} \sum_{j=1}^{n} (\mathbf{v}_i^{\diamond} \odot \mathbf{v}_j^{\diamond})\\
		&= \sum_{i=1}^{n} \sum_{j=i+1}^{n} (\mathbf{v}_i^{\diamond} \odot \mathbf{v}_j^{\diamond} + \mathbf{v}_j^{\diamond} \odot \mathbf{v}_i^{\diamond}),
\end{split}
\end{equation}
where $\odot$ denotes the \textbf{\textit{element-wise Hamilton product}} we define for two $d$-dimensional quaternion vectors. Take $\mathbf{v}_i^{\diamond} \odot \mathbf{v}_j^{\diamond}$ as an example, then:
\begin{equation}\label{eq:product_3}
\begin{split}
	\mathbf{v}_i^{\diamond} \odot \mathbf{v}_j^{\diamond} & = (\mathbf{r}_i\!\circ\! \mathbf{r}_j - \mathbf{a}_i\!\circ\! \mathbf{a}_j - \mathbf{b}_i\!\circ\! \mathbf{b}_j - \mathbf{c}_i\!\circ\! \mathbf{c}_j)\\
	& + (\mathbf{r}_i\!\circ\! \mathbf{a}_j + \mathbf{a}_i\!\circ\! \mathbf{r}_j + \mathbf{b}_i\!\circ\! \mathbf{c}_j - \mathbf{c}_i\!\circ\! \mathbf{b}_j)\mathbb{I}\\
	& + (\mathbf{r}_i\!\circ\! \mathbf{b}_j - \mathbf{a}_i\!\circ\! \mathbf{c}_j + \mathbf{b}_i\!\circ\! \mathbf{r}_j + \mathbf{c}_i\!\circ\! \mathbf{a}_j)\mathbb{J}\\
	& + (\mathbf{r}_i\!\circ\! \mathbf{c}_j + \mathbf{a}_i\!\circ\! \mathbf{b}_j - \mathbf{b}_i\!\circ\! \mathbf{a}_j + \mathbf{c}_i\!\circ\! \mathbf{r}_j)\mathbb{K},\\
\end{split}
\end{equation}
where $\circ$ represents the element-wise multiplication between two real-valued vectors. Similar to the inner Hamilton product, the element-wise Hamilton product $\odot$ also allows for the cross-channel information exchange between two quaternion embedding vectors as shown in Figure \ref{Figure:Struct}, where the quaternion cores interact with each other through element-wise product rather than inner product. For $\mathbf{v}^{\diamond}_i$ and $\mathbf{v}^{\diamond}_j$, the result of their element-wise Hamilton product is still a $d$-dimensional quaternion vector. Thus, compared with inner Hamilton product, element-wise Hamilton product has a lower compression rate when encoding pairwise feature interactions, but it preserves substantially more interaction contexts and makes Eq.(\ref{eq:pooling}) an ideal pooling operation for the subsequent quaternion-valued neural network to learn expressive latent representations.

\subsection{Quaternion Feed-Forward Network}
With the quaternion asymmetrical interaction pooling, all pairwise feature interactions are summarized in a unified representation $\widetilde{\mathbf{v}}^{\diamond}$. However, it is still a linear computation process. To further model the complex and non-linear interactions between different latent dimensions, we stack a quaternion-valued $l$-layer residual feed-forward network upon the interaction pooling layer:
\begin{equation}\label{eq:QFFN}
\begin{split}
	& \widetilde{\mathbf{h}}_1^{\diamond} = \widetilde{\mathbf{v}}^{\diamond} + \sigma(\mathbf{W}_1^{\diamond} \times \widetilde{\mathbf{v}}^{\diamond} + \mathbf{b}_1^{\diamond}),\\
	& \widetilde{\mathbf{h}}_2^{\diamond} = \widetilde{\mathbf{h}}_1^{\diamond} + \sigma(\mathbf{W}_2^{\diamond} \times \widetilde{\mathbf{h}}_1^{\diamond} + \mathbf{b}_2^{\diamond}),\\
	& \hspace{1.7cm} \cdots \\
	& \widetilde{\mathbf{h}}_l^{\diamond} = \widetilde{\mathbf{h}}_{l-1}^{\diamond} + \sigma(\mathbf{W}_l^{\diamond} \times \widetilde{\mathbf{h}}_{l-1}^{\diamond} + \mathbf{b}_l^{\diamond}),
\end{split}
\end{equation}
where $\mathbf{W}^{\diamond} \!\! = \!\! \mathbf{R}_w \!+\! \mathbf{A}_w\mathbb{I} \!+\! \mathbf{B}_w\mathbb{J} \!+\! \mathbf{C}_w\mathbb{K} \in \mathbb{H}^{d\times d}$ is the quaternion weight matrix, $\widetilde{\mathbf{h}}^{\diamond}$, $\mathbf{b}^{\diamond} \in \mathbb{H}^{d}$ are respectively the latent representation and bias in each layer, while $\sigma(\cdot)$ is the non-linear activation function. In the following, we introduce the three key components in the quaternion feed-forward network.

\textbf{Split Activation Function.} Activation functions plays a crucial role in modelling nonlinearity. Though there are few attempts to devise fully quaternion-valued activation functions by extending real-valued functions (e.g., tanh function) to the hypercomplex domain \cite{ujang2011quaternion,xia2014quaternion}, the quaternion-valued activation functions need a careful training phase due to an important number of singularities that can drastically alter a network's performance \cite{parcollet2019survey,parcollet2019quaternion_RNN}. As such, we adopt split activation for quaternion-valued neurons, which is widely adopted in  \cite{parcollet2019quaternion_RNN,tay2019lightweight,parcollet2019quaternion_CNN} due to its higher stability and simpler computation process. Specifically, for layer $l'\leq l$, let $\mathbf{h}_{l'}^{\diamond} = \mathbf{r}_h + \mathbf{a}_h\mathbb{I} + \mathbf{b}_h\mathbb{J} + \mathbf{c}_h\mathbb{K} = \mathbf{W}_{l'}^{\diamond} \times \widetilde{\mathbf{h}}_{l'-1}^{\diamond} + \mathbf{b}_{l'}^{\diamond}$, then $\sigma(\mathbf{h}_{l'}^{\diamond})$ is defined as:
\begin{align}\label{eq:split_activation}
		\sigma(\mathbf{h}^{\diamond}_{l'}) & = \sigma(\mathbf{r}_h + \mathbf{a}_h\mathbb{I} + \mathbf{b}_h\mathbb{J} + \mathbf{c}_h\mathbb{K})\\
	& = \!\textnormal{ReLU}(\mathbf{r}_h) \!+\! \textnormal{ReLU}(\mathbf{a}_h)\mathbb{I} \!+\! \textnormal{ReLU}(\mathbf{b}_h)\mathbb{J} \!+\! \textnormal{ReLU}(\mathbf{c}_h)\mathbb{K},\nonumber
\end{align}
where $\textnormal{ReLU}(\cdot)$ is the rectified linear unit we have chosen for nonlinear activation. With the split activation, the quaternion feed-forward network is able to simultaneously learn the latent representation of feature interactions through four separate channels (i.e., 1, $\mathbb{I}$, $\mathbb{J}$, $\mathbb{K}$), making the learned $\mathbf{h}^{\diamond}_l$ carry more intricate semantics about feature interactions.

\textbf{Residual Connections.}
The core idea behind residual networks is to propagate low-layer features to higher layers by residual connection \cite{he2016deep}. By combining low-layer interaction features with the high-layer representations computed by the feed-forward network, the residual connections essentially allow the model to easily propagate low-layer features to the final layer, which can help the model enhance its expressiveness using different information learned hierarchically. Intuitively, in our shared quaternion feed-forward network, to generate a comprehensive representation for feature interactions in each view, the $l'$-th layer iteratively fine-tunes the representation learned by the $(l'-1)$-th layer (i.e., $\widetilde{\mathbf{h}}^{\diamond}_{l'-1}$) by adding a learned residual, which corresponds to the second term in Eq.(\ref{eq:QFFN}).

\textbf{Layer Dropout.}
To prevent QNFM from overfitting the training data, we adopt dropout \cite{srivastava2014dropout} on all the layers of the quaternion feed-forward network as a regularization strategy. In short, we randomly drop the neurons with the ratio of $\rho \in (0, 1)$ during training. Hence, dropout can be viewed as a form of ensemble learning which includes numerous models that share parameters \cite{warde2013empirical}. It is worth mentioning that all the neurons are used when testing, which can be seen as a model averaging operation \cite{srivastava2014dropout} in ensemble learning.

\subsection{Comprehending Quaternion Feed-Forward Network} 
In Eq.(\ref{eq:QFFN}), the standard form of Hamilton product is used between the quaternion weight matrix and the quaternion vector. Recall that we can view the pooled quaternion vector $\widetilde{\mathbf{v}}^{\diamond}$ as a 4-channel representation containing four quaternion cores. As discussed in \cite{gaudet2018deep}, quaternion-valued matrix-vector multiplications is essentially mapping the original quaternion vector input to a new one by collectively using four sets of weights from the $1$, $\mathbb{I}$, $\mathbb{J}$, $\mathbb{K}$ channels, as illustrated in Figure~\ref{Figure:Struct}. Therefore, in the $l'$-th layer, $\mathbf{W}_{l'}^{\diamond} \times \widetilde{\mathbf{h}}_{l'-1}^{\diamond}$ produces a $d$-dimensional vector, where every single channel consists of a unique linear combination of each information channel in $\mathbf{W}_l'^{\diamond}$ and $\widetilde{\mathbf{h}}_{l'-1}^{\diamond}$. This comes from the properties of the Hamilton product, and is forcing each channel of the weight to interact with each channel of the input. Coupled with the bias terms, the split activation function and the residual connections, the quaternion feed-forward network is able to learn descriminative representations of subtle feature interactions via repetitive information exchange among the $1$, $\mathbb{I}$, $\mathbb{J}$, $\mathbb{K}$ channels.

\subsection{Output Layer}
The final vector representation $\mathbf{h}^{\diamond}_l$ is utilized to compute the quaternion output for the factorized feature interaction component via the inner Hamilton product of $\mathbf{h}^{\diamond}_l$ and weight $\mathbf{p}^{\diamond} \in \mathbb{H}^{d}$:
\begin{equation}\label{eq:output0}
	\phi'(\mathcal{V}_x^{\diamond}) = f_{\mathbb{H}\mapsto \mathbb{R}}(\mathbf{p}^{\diamond} \otimes \mathbf{h}_l^{\diamond}),
\end{equation}
where $f_{\mathbb{H}\rightarrow \mathbb{R}}(\cdot)$ is the same mapping function as defined in Eq.(\ref{eq:QFM_pred}). Lastly, we summarize the entire prediction result of QNFM as:
\begin{equation}\label{eq:output1}
	\widehat{y}= w_0 + \sum_{i=1}^{n}w_i x_i + f_{\mathbb{H}\mapsto \mathbb{R}}(\mathbf{p}^{\diamond} \otimes \mathbf{h}_l^{\diamond}).
\end{equation}

\section{Model Learning and Complexity Analysis}

\subsection{Model Learning}\label{sec:opt}
\textbf{Loss Function.}
As the scopes of both the input and output are not restricted, QFM and QNFM are flexible and versatile models which can be adopted for different tasks. In this paper, we focus on click-through rate (CTR) prediction, which is one of the most popular applications for FM-based models in recent years \cite{lian2018xdeepfm,juan2016field,shan2016deep,cheng2016wide,guo2017deepfm}. Given an arbitrary user and features like her/his profile or previously visited links (e.g., web pages or advertisements), CTR prediction aims to predict the possibility of clicking through a given link. To optimize QFM and QNFM towards the task goal, we enforce the model to output a probability of observing a $(user, link)$ instance. By replacing $(user, link)$ with the notion $(u,v)$, we quantify the prediction error with log loss, which is a special case of the cross-entropy:
\begin{equation}
	 L = - \!\!\!\!\! \sum_{(u_i, v_j)\in \mathcal{S}}{\!\!\! \Big{(} y_{ij}\log\beta(\widehat{y}_{ij}) \!+\! (1 \!-\! y_{ij})\log(1 \!-\! \beta(\widehat{y}_{ij})) \!\Big{)}},
\end{equation}
where $\beta(\cdot)$ is the sigmoid function to enable our model's classification capability such that $\beta(\widehat{y}_{ij}) \in (0,1)$ represents the possibility that user $u_i$ will click through link $v_j$. $\mathcal{S}$ is the set of ground truth $(u_i, v_j)$ pairs labelled by $y_{ij}\in\{0,1\}$. 

\textbf{Optimization Strategy.}
With the split activation in Eq.(\ref{eq:split_activation}) and real-valued loss $L$, we can efficiently apply stochastic gradient descent (SGD) algorithms to learn the model parameters by minimizing $L$. Hence, we leverage a mini-batch SGD-based algorithm, namely Adam~\cite{kingma2014adam} optimizer. For different tasks, we tune the hyperparameters using grid search. Specifically, the latent dimension (i.e., factorization factor) $d$ is searched in $\{4,8,16,32,64\}$; the depth of the quaternion feed-forward network $l$ is searched in $\{1,2,3,4,5\}$; and the dropout ratio $\rho$ is searched in $\{0.1,0.2,0.3,0.4,0.5\}$. We will further discuss the impact of these key hyperparameters to the prediction performance of QFM/QNFM in Section \ref{sec:sensitivity}. In addition, we set the batch size to 512 according to device capacity and the learning rate to $1\!\times\!10^{-5}$. The early stopping strategy is adopted, where the training process is terminated if the log loss on validation data keeps increasing for three consecutive training epochs.

\subsection{Space Complexity Analysis}\label{sec:space_complexity}
Since a major benefit of utilizing quaternion representations in FM-based models is the avoidance of excessive parameters, we analyze the space complexity of QFM and QNFM in detail. 
First, as pointed out in Section \ref{sec:rationale}, a $d$-dimensional quaternion representation can be viewed as an unravelled $4d$-dimensional real-valued vector. So, we should note that at the bit level, a $d$-dimensional quaternion vector has the same representation capacity as a vector in $\mathbb{R}^{4d}$, and will bring an identical number of $4d$ neurons in neural networks \cite{tay2019lightweight,parcollet2019quaternion_RNN}. In the same vein, our quaternion embedding scheme in Eq.(\ref{eq:embedding}) does not yield additional parameters as it just vertically splits a full embedding matrix in real-valued FM-models into 4 even partitions. Hence, we exclude the embedding layer in this discussion since it is mandatory in all FM-based models.
Also, we should recall that all the learnable parameters in the plain FM (i.e., Eq.(\ref{eq:FM})) apart from feature embeddings are the global bias $w_0$ and feature weights $\{w_i\}_{i=1}^{n}$, bringing only $n+1$ parameters in total.

\textbf{Space Complexity of QFM.} 
In QFM, we keep the original linear regression terms as in FM and only make changes to the feature interaction scheme. As Eq.(\ref{eq:QFM}) suggests, the quaternion-valued feature interaction function does not introduce any additional parameters to learn. Thus, the number of trainable parameters in QFM is identical to the lightweight FM.

\textbf{Space Complexity of QNFM.}
We start with the quaternion feed-forward network in QNFM, where its advantage of parameter reduction can be revealed in the comparison with real-valued deep networks. 
To be consistent, we benchmark the parameter size with $4d$ neurons in a neural network, which corresponds to $\mathbb{H}^d$ in the hypercomplex space as per our discussion. In each feed-forward layer of QNFM, the trainable parameters are the quaternion cores in weight $\mathbf{W}^{\diamond}$ and bias $\mathbf{b}^{\diamond}$, consuming $4d^2$ and $4d$ parameters per layer, respectively. In contrast, though an equivalent single-layer, real-valued neural network with $4d$ neurons has a bias vector of the same size, its $4d\times 4d$ weight matrix yields $16d^2$ parameters in total. To explain where the much lower parameter consumption comes from, we take $d=1$ as an example in the $l'$-layer of Eq.(\ref{eq:QFFN}). Intuitively, if we set aside the quaternion unit basis $1$, $\mathbb{I}$, $\mathbb{J}$ and $\mathbb{K}$, the Hamilton product $\mathbf{W}^{\diamond}_{l'} \times \mathbf{h}^{\diamond}_{l'}$ can be approximated as:
\begin{equation}
	\mathbf{W}^{\diamond}_{l'} \times \mathbf{h}^{\diamond}_{l'} \sim 
	\begin{bmatrix}
		r_w & -a_w & -b_w & -c_w\\
		a_w & r_w & -c_w & b_w\\
		b_w	& c_w & r_w & -a_w\\
		c_w & -b_w & a_w & r_w\\
	\end{bmatrix}
	\times 
	\begin{bmatrix}
		r_h\\
		a_h\\
		b_h\\
		c_h\\
	\end{bmatrix},
\end{equation}
where there are only 4 distinct parameter variables (i.e., quaternion cores $r_w$, $a_w$, $b_w$ and $c_w$) in the weight matrix, leading to as few as 4 degrees of freedom in trainable parameters. However, in every real-valued feed-forward operation, all the 16 elements in the weight matrix are different parameter variables \cite{tay2019lightweight}. Hence, ignoring the small bias vector, our quaternion feed-forward operation achieves a $75\%$ reduction in parameterization. Meanwhile, similar to QFM, the size of model parameters in the first two regression terms of Eq.(\ref{eq:QFFN}) is $n+1$. In Eq.(\ref{eq:pooling}) which is our proposed asymmetrical quaternion interaction pooling layer, there are none learnable parameters. As the projection weight $\mathbf{p}^{\diamond} \in \mathbb{H}^{d}$ in the output layer has $4d$ parameters, the total parameter size of QNFM is $n+1+ l\times(4d^2 + 4d) + 4d$. Considering the $n+1$ parameters in the plain FM, the extra space complexity introduced by QNFM on top of FM is $O(d^2l+dl+d)$. On the contrary, an equivalently structured NFM \cite{he2017neural} which is one of the simplest DNN-based FM variants, has the extra space complexity of $O(4d^2l+dl+d)$ versus FM. As $l\ll d$ in practice, our QNFM is a substantially more compact model than its real-valued counterparts, thus being highly memory-efficient.

\subsection{Time Complexity Analysis}
Excluding the embedding operation that is standard in all FM-based models, the major computational cost of our model is exerted by the feature interaction functions $\phi(\cdot)$ and $\phi'(\cdot)$ respectively in QFM and QNFM. In what follows, we prove that their time complexity is linearly associated to only the number of instances in a dataset.

\textbf{Time Complexity of QFM.}
The core computational step in $\phi(\cdot)$ is the feature interaction scheme based on inner Hamilton product defined in Eq.(\ref{eq:twoway_interaction}), which has the time complexity of $O(n^2d)$ for each data instance. Because we only account for non-zero features (i.e., $\mathcal{V}_x$), the actual time complexity is $O(|\mathcal{V}_x|^2d)$ where $|\mathcal{V}_x| \ll n$. As the number of non-zero features (i.e., feature fields) in $\mathcal{V}_x$ is constant and commonly small in CTR prediction datasets (see Section \ref{sec:datasets}), and the latent dimension $d$ is also fixed, QFM has linear time complexity w.r.t. the scale of the data.

\textbf{Time Complexity of QNFM.}
In QNFM, the time complexity of $\phi'(\cdot)$ is mainly brought by the asymmetrical quaternion interaction pooling, quaternion feed-forward network, and the projection operation within the output layer, which respectively takes $O(|\mathcal{V}_x|^2d)$, $O(d^2)$ and $O(d)$ time to compute. Hence, the total time complexity tallies up to $O(|\mathcal{V}_x|^2d + d^2 + d)$ per instance, which is also linearly related to the data size.

\section{Related Work}\label{sec:related}

\subsection{Feature Interaction-based Predictive Analytics}

When performing sparse predictive analytics,
 learning the context of interactions among observed features is an indispensable means for machine learning models to achieve optimal results \cite{shan2016deep}. On this occasion, factorization machine (FM) \cite{rendle2010factorization} is a well-established model that captures the second-order feature interactions for predictive analytics. Following FM, a series of its linear variants are proposed, such as field-aware FM \cite{juan2016field}, higher-order FM \cite{blondel2016higher} and importance-aware FM \cite{oentaryo2014predicting}. However, as stated in many literatures \cite{qu2016product,lian2018xdeepfm,he2017neural,chen2020sequence}, these FM extensions are constrained by their limited expressiveness when mining high-order interaction patterns and learning discriminative feature representations.

With the fruitful advances in deep neural networks (DNNs) \cite{lecun2015deep}, research that incorporates DNNs into the feature interaction paradigm has gained state-of-the-art performance in a considerable amount of sparse prediction tasks \cite{zhang2016deep,he2017neural,qu2016product,lian2018xdeepfm}. 
Generally, there are two main themes of leveraging DNNs to learn feature interactions, which are developing ``deep'' and/or ``wide'' models \cite{chen2020sequence}, respectively. In a deep structure,  multi-layer DNNs are commonly stacked upon linear components to exhaustively extract useful information from feature interactions, e.g., the residual network in DeepCrossing \cite{shan2016deep}, the convolutional network in FGCNN \cite{liu2019feature}, and the compressed interaction network in xDeepFM \cite{lian2017practical}. To build wide models, multiple types of feature interactions from varied domains (if available) are taken into account, e.g., separately modelling user logs and texts with CoFM \cite{hong2013co}, differentiating user and item contexts in ENSFM \cite{chen2020efficient}, and applying multi-view learning to dynamic and static features in SeqFM \cite{chen2020sequence}. It is worth mentioning that, in recent hybrid models, by fusing shallow low-order (e.g., linear regression) output with dense high-order (e.g., DNN) output, deep and wide structures can be united to offer richer prediction signals, such as Wide\&Deep \cite{cheng2016wide}, DeepFM \cite{guo2017deepfm}, DCN \cite{wang2017deep}, and AutoInt \cite{song2019autoint}. 

As pointed out in Section \ref{sec:intro}, compared with the plain FM, though DNN-based methods are more advantageous in feature interaction modelling, the heavy parameterization is a mandatory prerequisite for those high-performance models. Consequently, the excessive parameter consumption of aforementioned deep models creates a severe bottleneck for efficient deployment and training in the real world. 

\subsection{Quaternion Representations and Neural Networks}
In a nutshell, DNNs aim to learn discriminative latent representations from the data for downstream tasks. Among various neural architectures, the majority of works have mainly explored the usage of real-valued representations. Recently, an emerging research interest in complex \cite{danihelka2016associative,arjovsky2016unitary} and hypercomplex (i.e., quaternion) representations \cite{gaudet2018deep,parcollet2019quaternion_CNN,parcollet2019quaternion_RNN} has arisen. For example, Parcollet \textit{et al.} successfully make the first attempt to devise a quaternion multi-layer perceptron for language understanding \cite{parcollet2016quaternion}. For image classification, two highly effective quaternion convolutional neural networks are proposed in \cite{gaudet2018deep} and \cite{zhu2018quaternion}, along with fundamental tools like quaternion-valued convolution operation, batch normalization, and initialization. In the pursuit of maximum performance, both recurrent neural and convolutional networks are endowed with quaternion representations and achieve dominating results in different tasks, such as speech recognition \cite{parcollet2019quaternion_RNN}, sound detection \cite{comminiello2019quaternion}, and image conversion \cite{parcollet2019quaternion_CNN}.

Apart from quaternion neural networks, quaternion representations also witness state-of-the-art performance in other areas like knowledge graph embedding \cite{zhang2019quaternion} and collaborative filtering \cite{zhang2019quaternion_CF}. Notably, progress on quaternion representations for deep learning is still in its infancy \cite{tay2019lightweight}, making its immense potential largely unexplored. In short, the benefit from quaternion representations can be attributed to several factors. Firstly, quaternions are intuitively linked to associative composition \cite{tay2019lightweight}, leading to higher representation capacity and generalizability than real-valued counterparts \cite{zhang2019quaternion_CF}. Also, the hypercomplex Hamilton product provides a greater extent of expressiveness with a four-fold increase in interactions between real and imaginary components. Secondly, the asymmetry of Hamilton product has also demonstrated promising advantages in learning entity relationships in knowledge graphs \cite{zhang2019quaternion} and inferring user-item preferences in recommender systems \cite{zhang2019quaternion_CF}. Thirdly, in the case of quaternion representations, due to the $75\%$ parameter reduction in the Hamilton product, quaternion neural networks also enjoy a significant reduction in parameter size \cite{parcollet2019quaternion_RNN,tay2019lightweight,parcollet2019quaternion_CNN}, and can alleviate the risk of overfitting brought by too many degrees of freedom in learnable parameters \cite{zhu2018quaternion}.

\subsection{Network Compression}
Our work is related to, but highly distinct from the research on network compression, among which two major representatives are network pruning \cite{han2015learning} and quantization \cite{zhao2019improving}. Network pruning reduces the weight parameters needed in neural models by cutting off unnecessary connections between neurons, while network quantization maps all continuous weight parameters into a fixed set of discrete values to minimize the total bits needed to store them \cite{han2015deep}. However, those pruning and quantization methods are specifically designed to compress a well-trained large neural network, and this is non-comparable to our motivation of designing and training a lightweight prediction model from scratch. Furthermore, such network compression paradigms have not yet been utilized in FM-based models for predictive analytics, leaving it a promising direction for future research.

\section{Experiments}\label{sec:exp}
In this section, we first outline the evaluation settings regarding the CTR prediction task. Then, we conduct experiments to evaluate QFM and QNFM in terms of both effectiveness and efficiency. In particular, we aim to answer the following research questions (RQs) via experiments:
\begin{description}
	\item[\textbf{RQ1:}] How effectively can QFM and QNFM perform sparse predictive analytics compared with state-of-the-art FM-based models?
	\item[\textbf{RQ2:}] How is the parameter efficiency of our models?
	\item[\textbf{RQ3:}] How do the hyperparameters affect the performance of QFM and QNFM in CTR prediction?
	\item[\textbf{RQ4:}] How do QFM and QNFM benefit from each component of the proposed model structure?
	\item[\textbf{RQ5:}] How is the training efficiency and scalability of both models when handling large-scale data?
\end{description}

\subsection{Datasets}\label{sec:datasets}
To validate the performance of our proposed models in terms of CTR prediction accuracy, we adopt four real-world datasets. All datasets used in our experiment are in large scale and publicly available. The primary statistics are shown in Table~\ref{table:Dataset}, and their properties are introduced below.
\begin{itemize}
	\item \textbf{Trivago:} This dataset is from the 2019 ACM RecSys Challenge\footnote{http://www.recsyschallenge.com/2019/}. It is a web search dataset consisting of users' visiting logs (e.g., city of the user, device being used, etc.) on different hotel booking webpages.
	\item \textbf{Criteo:} This is an industry-level benchmark dataset\footnote{http://labs.criteo.com/2014/02/} for CTR prediction. As our largest dataset, it has more than 45 million users' clicking instances on displayed advertisements, with a mixture of 26 categorical feature fields and 13 numerical feature fields. 
	\item \textbf{Avazu:} This benchmark dataset\footnote{https://www.kaggle.com/c/avazu-ctr-prediction} is similar to Criteo in size. It records mobile users' behaviors including whether or not a displayed advertisement is clicked. 
\end{itemize}

\begin{table}[!t]
\small
\caption{Statistics of datasets in use.}
\vspace{-0.5cm}
\renewcommand{\arraystretch}{1.0}
\setlength\tabcolsep{4pt}
\center
  \begin{tabular}{c c c c}
    \toprule
    Dataset & \#Instance & \#Field& \#Feature (Sparse)\\
	\midrule
	Trivago & 2,464,024 & 30 & 1,331,752\\
	Criteo & 45,840,617 & 39 & 998,960\\
	Avazu & 40,428,967 & 23 & 1,544,488\\
    \bottomrule
\end{tabular}
\label{table:Dataset}
\vspace{-0.3cm}
\end{table}

\begin{table*}[t]
\small
\caption{CTR prediction results. Numbers in bold face are the best results for corresponding metrics.}
\vspace{-0.3cm}
\centering
\renewcommand{\arraystretch}{1.0}
\setlength\tabcolsep{10pt}
  \begin{tabular}{|c|c|c|c||c|c|c||c|c|c|}
    \hline
    \multirow{2}{*}{Method} & \multicolumn{3}{c||}{Trivago}  & \multicolumn{3}{c||}{Criteo}  & \multicolumn{3}{c|}{Avazu}\\
    \cline{2-10}
    & AUC & LE & RMSE & AUC & LE & RMSE & AUC & LE & RMSE\\
    \hline
    FM \cite{rendle2010factorization}& 0.8186 & 0.5334 & 0.4191 & 0.7893 & 0.4650 & 0.3939 & 0.7712 & 0.3859 & 0.3489 \\
    DeepCrossing \cite{shan2016deep}& 0.8510 & 0.4408 & 0.3993 & 0.8007 & 0.4508 & 0.3869 & 0.7624 & 0.3910 & 0.3503 \\
    NFM\cite{he2017neural}& 0.8487 & 0.4427 & 0.3834 & 0.7970 & 0.4557 & 0.3836 & 0.7706 & 0.3855 & 0.3485 \\
    AFM\cite{xiao2017attentional}& 0.8363 & 0.4652 & 0.4037 & 0.7931 & 0.4594 & 0.3915 & 0.7722 & 0.3853 & 0.3484 \\
    DCN\cite{wang2017deep}& 0.8573 & 0.4449 & 0.3849 & 0.8010 & 0.4513 & 0.3854  & 0.7683 & 0.3873 & 0.3501\\
    xDeepFM\cite{lian2018xdeepfm}& 0.8703 & 0.4372 & 0.3823 & 0.8029 & 0.4487 & 0.3815  & 0.7746 & 0.3842 & 0.3479\\
    AutoInt\cite{song2019autoint}& 0.8653 & 0.4394 & 0.3851 & 0.8055 & 0.4469 & 0.3812  & 0.7747 & 0.3837 & 0.3474\\
    HFM\cite{tay2019holographic} & 0.8482 & 0.4441 & 0.3849 & 0.8014 & 0.4499 & 0.3817 & 0.7670 & 0.3869 & 0.3490 \\
    HFM+\cite{tay2019holographic} & 0.8664 & 0.4392 & 0.3796 & 0.8035 & 0.4476 & 0.3810 & 0.7665 & 0.3871 & 0.3485 \\ 
    \hline
     \!\textbf{QFM}& 0.8543 & 0.4696 & 0.3928 & 0.8025 & 0.4492 & 0.3812 & 0.7739 & 0.3847 & 0.3475\\
     \textbf{QNFM}& \textbf{0.8838} & \textbf{0.4323} & \textbf{0.3740} & \textbf{0.8062} & \textbf{0.4454} & \textbf{0.3803} & \textbf{0.7758} & \textbf{0.3823} & \textbf{0.3469}\\
     \hline
    \end{tabular}
\label{table:performance}
\vspace{-0.4cm}
\end{table*}

\subsection{Baseline Methods}
We briefly introduce the baselines for comparison below.
\begin{itemize}
	\item \textbf{FM:} This is the original FM model \cite{rendle2010factorization} with proven effectiveness in many prediction tasks.
	\item \textbf{DeepCrossing:} It stacks multiple residual network blocks upon the concatenation layer for feature embeddings in order to learn deep cross features \cite{shan2016deep}. 
	\item \textbf{NFM:} The neural FM \cite{he2017neural} encodes all feature interactions via a multi-layer neural network coupled with a bit-wise bi-interaction pooling layer.
	\item \textbf{AFM:} The attentional FM \cite{xiao2017attentional} introduces an attention network to distinguish the importance of different pairwise feature interactions.
	\item \textbf{DCN:} The deep cross network \cite{wang2017deep} takes the outer product  of concatenated feature embeddings to explicitly model feature interaction.
	\item \textbf{xDeepFM:} It stands for the extreme deep FM \cite{lian2018xdeepfm} that has a compressed interaction network to model vector-wise feature interactions for CTR prediction.
	\item \textbf{AutoInt:} Having a multi-head self-attentive neural network as the core component \cite{song2019autoint}, it automatically learns the high-order interactions of input features.
	\item \textbf{HFM:} The holographic FM \cite{tay2019holographic} uses circular convolutions for feature interaction modelling. The idea behind HFM is similar to ours, i.e., maximizing the pairwise feature interaction modelling with no additional parameters upon the plain FM.
	\item \textbf{HFM+:} This is the DNN-enhanced version of HFM proposed by \cite{tay2019holographic}.
\end{itemize}

Notably, FM, AFM and HFM are for second-order feature interaction modelling, while all other baselines account for higher-order interactions. Dedicated to CTR prediction, DeepCrossing, xDeepFM, DCN and AutoInt are currently the state-of-the-art methods in this line of research.

\subsection{Evaluation Protocols}
For each dataset, we randomly split its instances with a ratio of 8:1:1 for training, validation, and test, respectively. We adopt two well-established evaluation metrics for CTR prediction \cite{shan2016deep,lian2018xdeepfm}, namely Area under the ROC Curve (AUC) and Log Error (LE, equivalent to cross-entropy in our case). Furthermore, as the output in this task is a probability that approximates the binary label $\{0,1\}$, we further adopt Root Mean Squared Error (RMSE) \cite{he2017neural,xiao2017attentional}. AUC measures the probability that a positive instance will be ranked higher than the negative one. It only takes into account the order of predicted instances and is insensitive to class imbalance problem. In contrast, LE and RMSE straightforwardly evaluate the distance between the predicted possibility and the true label for each instance. It is worth noting that, in the context of CTR prediction where service providers commonly have a large user base, an improvement on the aforementioned metrics at the \textbf{\textit{0.001-level}} is regarded significant \cite{song2019autoint,wang2017deep,li2020interpretable,guo2017deepfm}.

\subsection{Hyperparameter Settings}
To be consistent, we report the overall performance of our models with $d=64$ for QFM and a unified parameter set $\{d=64, l=1, \rho = 0.1\}$ for QNFM. Note that as described in Section \ref{sec:space_complexity}, a 64-dimensional quaternion representation in QFM/QNFM actually takes up 256 bits to represent all the information. Details on the effect of different hyperparameter settings will be given in Section \ref{sec:sensitivity}. For all baselines, we set the dimensions of both the embedding layer and hidden layer (if there is any) to 256 to ensure the same representation capacity as QFM/QNFM, and tune all remaining hyperparameters via grid search. Specifically, we adopt a 2-tier residual network in DeepCrossing, and the numbers of deep feature interaction layers in DCN, xDeepFM and AutoInt are all set to 3 while xDeepFM is also coupled with a 2-layer feed-forward network by default. For HFM+, we use its original setting of 3 feedforward layers. For both NFM and AFM, we implement a single-layer structure in their feed-forward network and attention network, respectively. 

\begin{figure*}[t!]
\centering
\begin{tabular}{cccc}
	\vspace{-0.2cm}\includegraphics[width=1.75in]{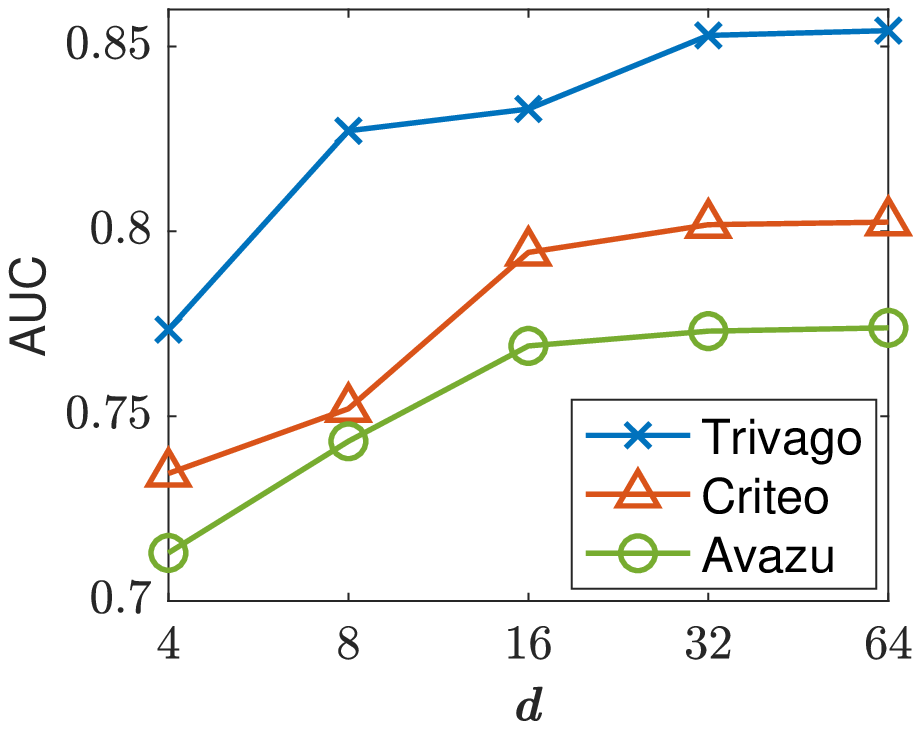}
	&\hspace{-0.6cm}\includegraphics[width=1.75in]{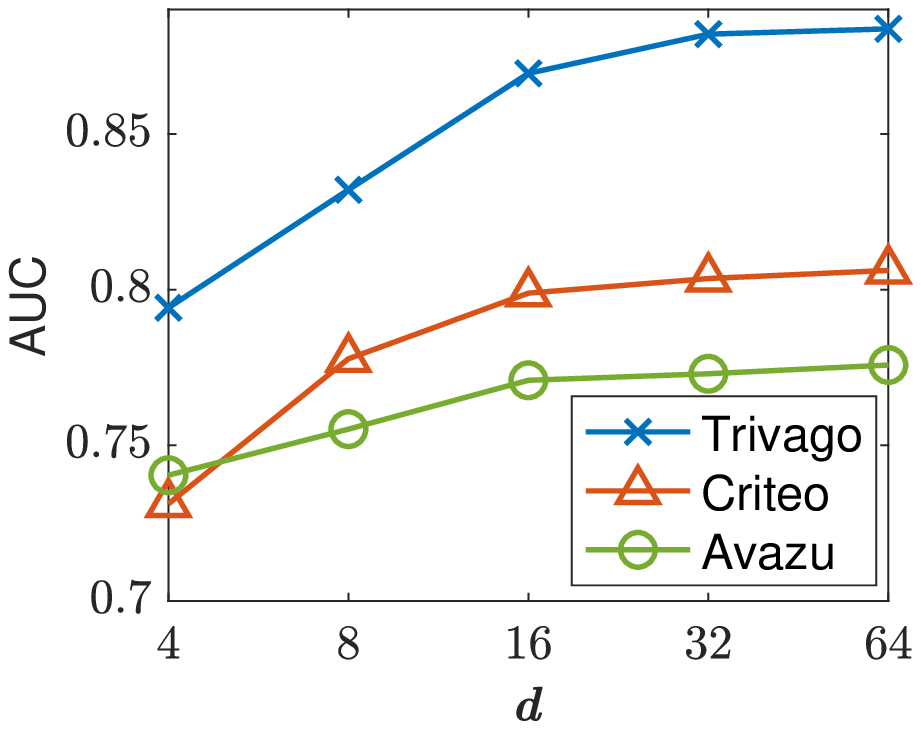}
	&\hspace{-0.6cm}\includegraphics[width=1.75in]{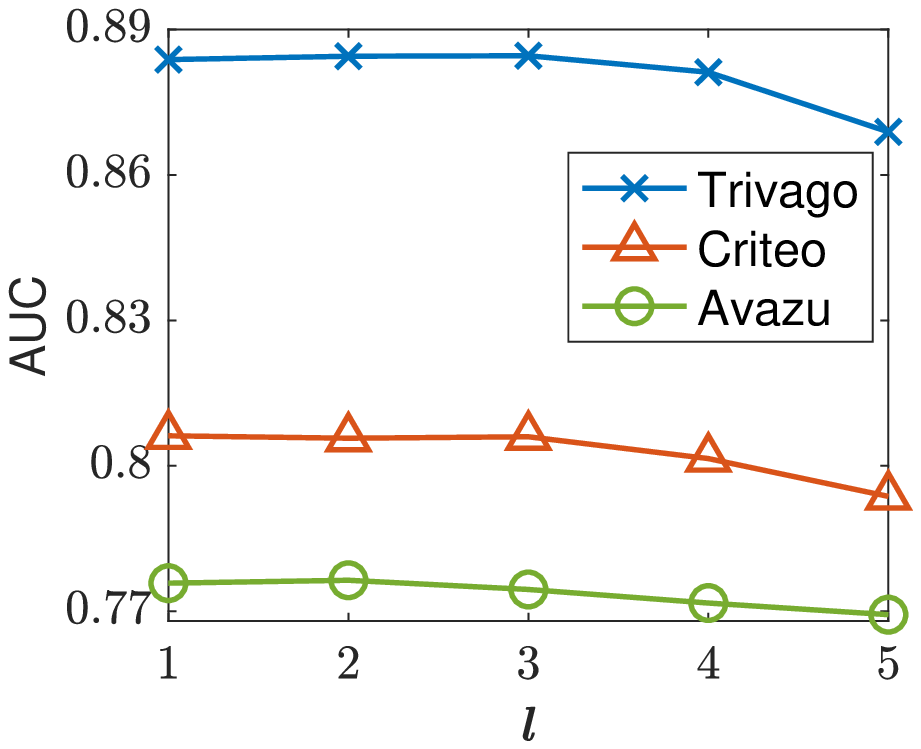}
	&\hspace{-0.6cm}\includegraphics[width=1.75in]{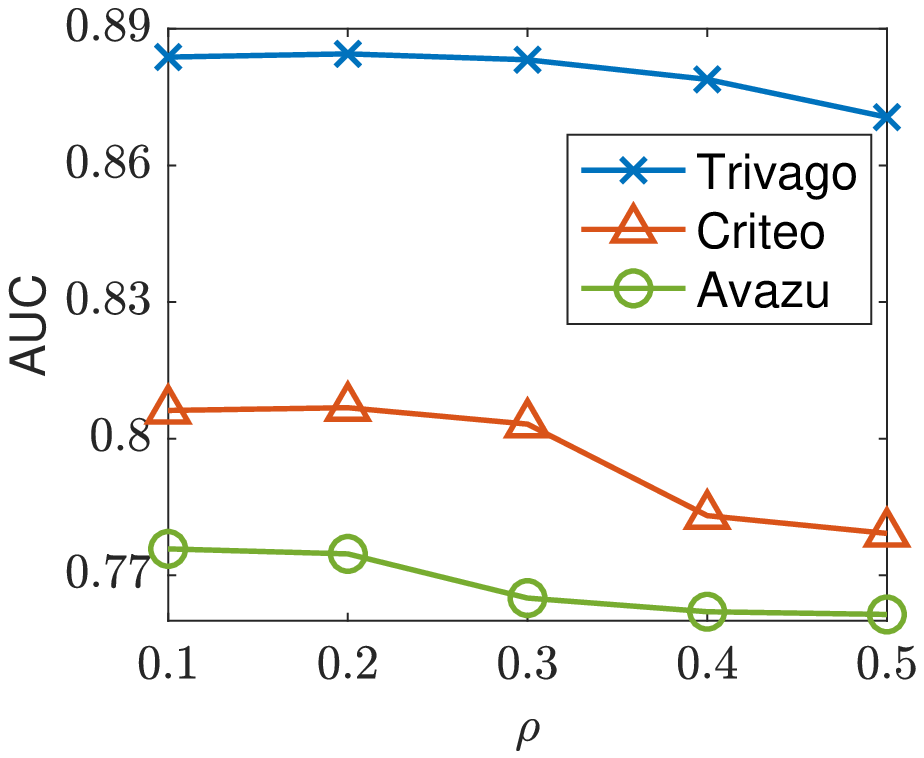}\\
	\footnotesize{(a) AUC of QFM w.r.t. $d$.}
	&\footnotesize{(b) AUC of QNFM w.r.t. $d$.}
	&\footnotesize{(c) AUC of QNFM w.r.t. $l$.}
	&\footnotesize{(d) AUC of QNFM w.r.t. $\rho$.}\\
	\vspace{-0.2cm}\includegraphics[width=1.75in]{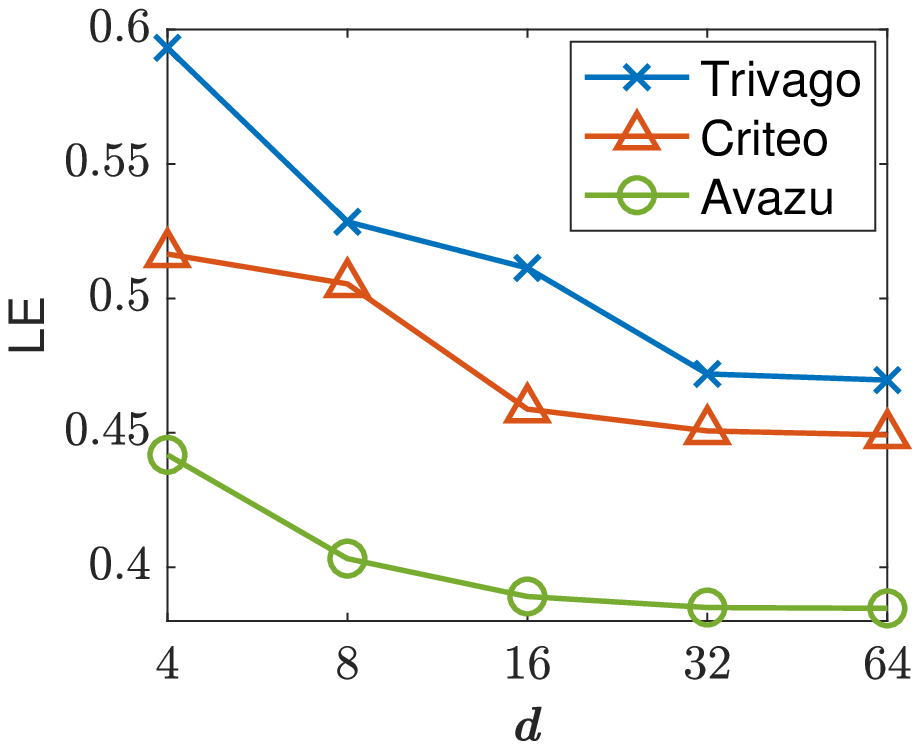}
	&\hspace{-0.6cm}\includegraphics[width=1.75in]{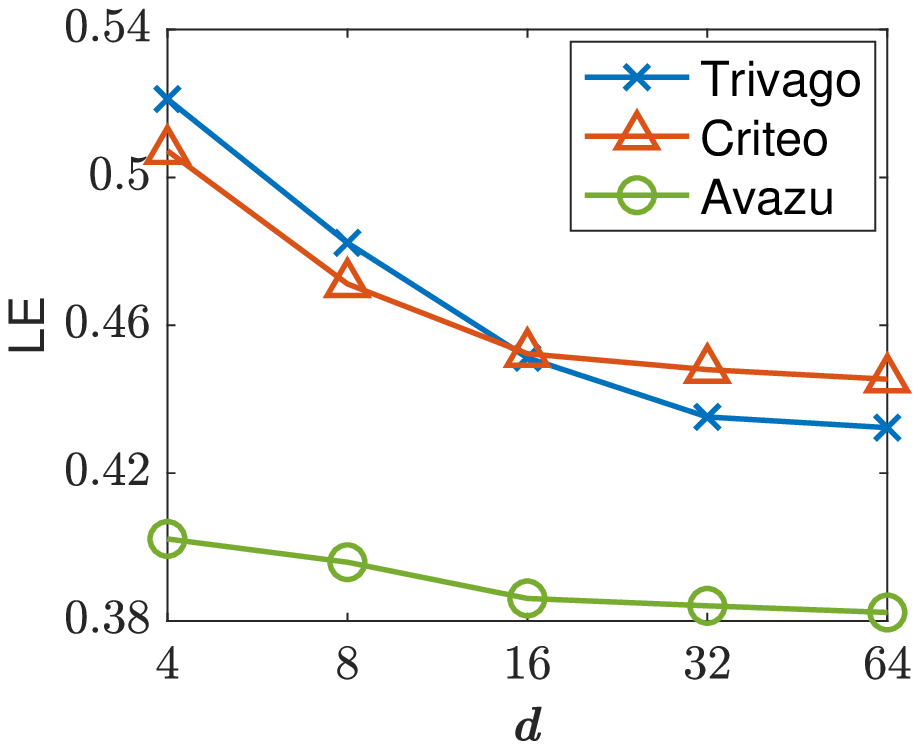}
	&\hspace{-0.6cm}\includegraphics[width=1.75in]{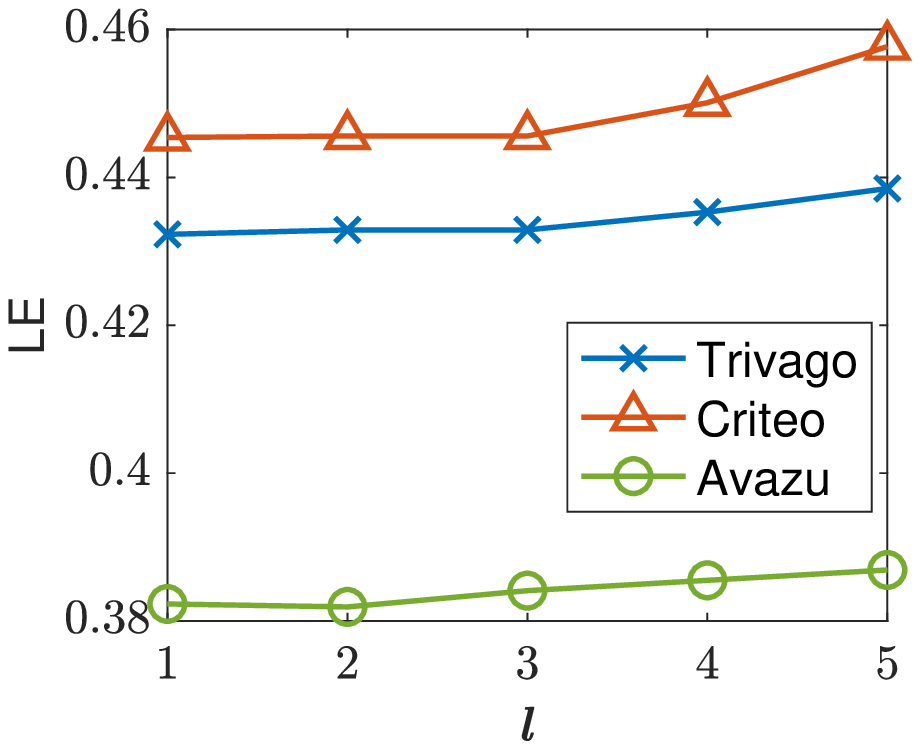}
	&\hspace{-0.6cm}\includegraphics[width=1.75in]{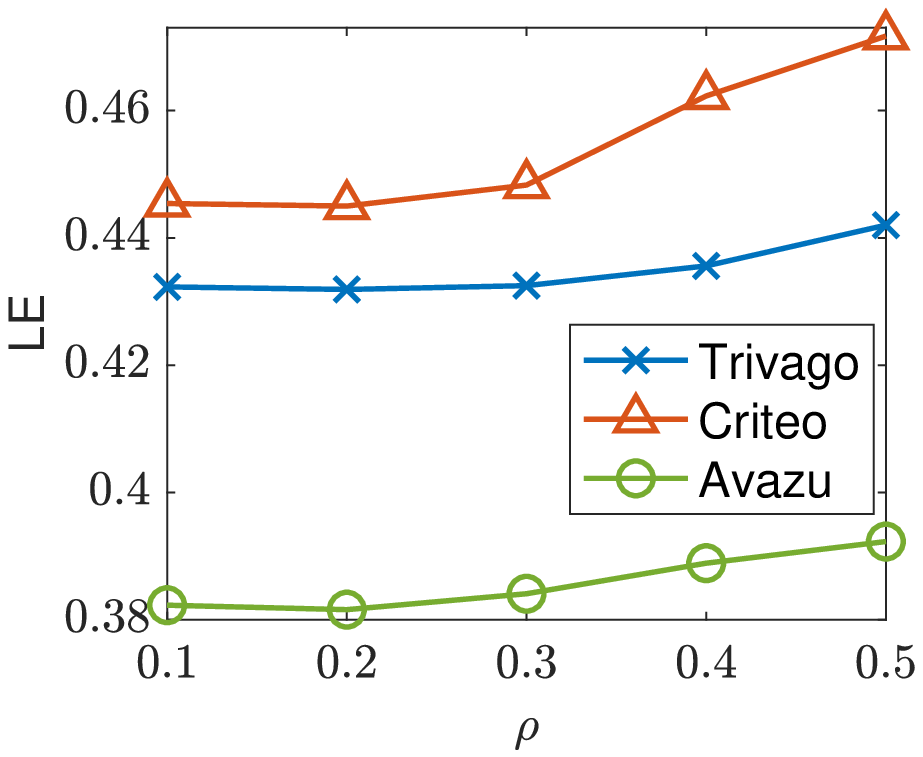}\\
	\footnotesize{(e) LE of QFM w.r.t. $d$.}
	&\footnotesize{(f) LE of QNFM w.r.t. $d$.}
	&\footnotesize{(g) LE of QNFM w.r.t. $l$.}
	&\footnotesize{(h) LE of QNFM w.r.t. $\rho$.}\\
		\end{tabular}
\vspace{-0.1cm}
\caption{Prediction performance w.r.t. $d$, $l$, and $\rho$. (a) and (e) correspond to QFM, while (b)-(d) and (f)-(h) correspond to QNFM.}
\label{Figure:paramsensitivity}
\vspace{-0.5cm}
\end{figure*}

\subsection{Effectiveness Analysis (RQ1)}\label{sec:effectiveness}
We showcase all models' performance on CTR prediction with Table \ref{table:performance}. Based on the experimental results, we draw the following observations.

First of all, our proposed QNFM, which is the neural network-enhanced version, consistently and significantly outperforms all the baseline methods. This verifies the efficacy of Hamilton product in modelling the intricate feature interactions. In particular, on the highly sparse Trivago dataset which has a similar total number of features as Avazu but 16 times less observed training instances, the quaternion-valued feature interaction scheme in QNFM can fully learn the contexts of interactions between available features, thus yielding $1.6\%$ performance gain on AUC over the second best model (i.e., xDeepFM). Meanwhile, as we will demonstrate in Section \ref{sec:efficiency}, QNFM achieves the best prediction results with substantially lower parameter consumption compared with all state-of-the-art baselines.

Secondly, as a relatively shallow and simple model, QFM shows strong competitiveness in terms of prediction performance. For example, on Criteo and Avazu, QFM outperforms the majority of all baselines except for xDeepFM and AutoInt. It also produces a significant improvement over the plain FM on all three datasets with an average performance increase of $2.13\%$, $5.22\%$, $3.30\%$ on AUC, LE, and RMSE, respectively.

Lastly, the results imply that deep neural network-based FM variants are generally more powerful in sparse predictive analytics. Though AFM utilizes a deep layer to compute the an attention weight for every pairwise feature interaction, the combinatorial effect of all feature interactions are still merged in a linear fashion, leading to inferior performance among other deep FM-based models like DCN. Another interesting finding is that straightforwardly using deep networks for high-order feature interaction modelling does not necessarily lead to better prediction performance, which is evidenced by the slightly lower AUC scores achieved by DeepCrossing and NFM on Avazu in comparison to FM and AFM.

\begin{table}[t]
\small
\caption{Comparison on extra parameters and AUC performance gain versus the plain FM. The results are based on the same embedding and latent dimension of 256 in $\mathbb{R}$, which is equivalent to $d=64$ in $\mathbb{H}$.}
\vspace{-0.3cm}
\centering
\renewcommand{\arraystretch}{1.05}
\setlength\tabcolsep{1.0pt}
  \begin{tabular}{|c|c c c||c c c|}
    \hline
    \multirow{3}{*}{Method} & \multicolumn{3}{c||}{Extra Parameter}  & \multicolumn{3}{c|}{AUC Performance}\\
    & \multicolumn{3}{c||}{Size vs. FM}  & \multicolumn{3}{c|}{Gain vs. FM}\\
    \cline{2-7}
    & Trivago & Criteo & Avazu & Trivago & Criteo & Avazu \\
    \hline
    DeepCrossing \cite{shan2016deep} & 2.0$\times\!$10\textsuperscript{6} & 2.6$\times\!$10\textsuperscript{6} & 1.5$\times\!$10\textsuperscript{6}   & 3.96\% & 1.44\% & -1.14\% \\
    NFM\cite{he2017neural}& 6.6$\times\!$10\textsuperscript{4} & 6.6$\times\!$10\textsuperscript{4} & 6.6$\times\!$10\textsuperscript{4} & 3.68\%  & 0.98\%  & -0.08\% \\
    AFM\cite{xiao2017attentional}& 6.6$\times\!$10\textsuperscript{4} & 6.6$\times\!$10\textsuperscript{4} & 6.6$\times\!$10\textsuperscript{4} & 2.16\% & 0.48\% & 0.13\% \\
    DCN\cite{wang2017deep} & 2.1$\times\!$10\textsuperscript{6} & 2.7$\times\!$10\textsuperscript{6} & 1.7$\times\!$10\textsuperscript{6} & 4.73\% & 1.48\% & -0.38\% \\
    xDeepFM\cite{lian2018xdeepfm} & 2.1$\times\!$10\textsuperscript{6} & 2.8$\times\!$10\textsuperscript{6}  & 1.6$\times\!$10\textsuperscript{6}  & 6.32\% & 1.72\% & 0.44\% \\
    AutoInt\cite{song2019autoint}& 1.0$\times\!$10\textsuperscript{6}& 1.0$\times\!$10\textsuperscript{6}  & 9.9$\times\!$10\textsuperscript{5}  & 5.70\% & 2.05\% & 0.45\% \\
    HFM\cite{tay2019holographic} & \textbf{0} & \textbf{0} & \textbf{0} & 3.62\% & 1.53\% & -0.54\% \\
    HFM+\cite{tay2019holographic} & 2.0$\times\!$10\textsuperscript{5} & 2.0$\times\!$10\textsuperscript{5} & 2.0$\times\!$10\textsuperscript{5} & 5.84\% & 1.80\% & -0.61\% \\ 
    \hline
     \textbf{QFM}& \textbf{0} & \textbf{0} & \textbf{0} & 4.36\% & 1.67\% & 0.35\% \\
     \textbf{QNFM}& 1.6$\times\!$10\textsuperscript{4} & 1.6$\times\!$10\textsuperscript{4} & 1.6$\times\!$10\textsuperscript{4} & \textbf{7.96\%} & \textbf{2.14\%} & \textbf{0.60\%} \\
     \hline
    \end{tabular}
\vspace{-0.6cm}
\label{table:param_efficiency}
\end{table}

\subsection{Parameter Efficiency Analysis (RQ2)}\label{sec:efficiency}
As we target on high-performance FM-based models that can minimize the parameter costs, we compare all models' parameter efficiency w.r.t. their performance. In Table \ref{table:param_efficiency}, we report the extra parameter size introduced by each model on top of the plain FM, as well as their performance gain on AUC versus FM. Apparently, it is obvious that DeepCrossing, DCN and xDeepFM are the most memory-consuming. However, apart from xDeepFM, the performance gain from DeepCrossing and DCN does not offset their excessive model sizes. For instance, though $1.5$ and $1.7$ millions of extra parameters are respectively brought by DeepCrossing and DCN on Avazu, they both fail to improve the performance of FM. On all three datasets, xDeepFM requests for an average of $0.77$ million additional parameters for every $1\%$ performance gain over the plain FM. 

In contrast to all baselines, this is where the proposed QFM and QNFM shine. On one hand, QFM obtains similar performance on Trivago as DCN and outperforms it on both Criteo and Avazu, but all the performance gain over the FM model is achieved \textbf{\textit{with no extra parameters}}. Compared with HFM which also has none extra parameters, QFM consistently outperforms it on all three datasets, proving that quaternion representations are more effective in modelling feature interactions. Hence, being as lightweight as the plain FM, QFM can produce highly competitive prediction results that are in line with or even better than multiple strong baselines. On the other hand, in most cases, the total number of extra parameters introduced by QNFM is two magnitudes lower than the heavily parameterized models including DeepCrossing, DCN, xDeepFM and AutoInt. Besides, unlike most of the baseline methods, an important property of QFM and QNFM is that the model size is invariant to the amount of feature fields, making both models highly compact on large-scale datasets. Compared with NFM and AFM using single-layer networks, QNFM aggressively reduces the parameter size by approximately 4 times with a significant performance boost. In addition, for QNFM, the average parameter cost per $1\%$ performance gain is just over $2,000$ on all datasets, which fully shows the superiority of learning feature interactions in the hypercomplex space. 

\subsection{Hyperparameter Sensitivity (RQ3)}\label{sec:sensitivity}
We answer RQ3 by investigating the performance fluctuations of QFM/QNFM with varied hyperparameters. Particularly, as mentioned in Section \ref{sec:opt}, we study the impact of latent dimension $d$ on both QFM and QNFM. We further investigate QNFM's sensitivity to the feed-forward network depth $l$ and the dropout ratio $\rho$. Based on the standard setting $\{d=64, l=1, \rho=0.1\}$ of QNFM, we vary the value of one hyperparameter while keeping the others unchanged, and record the new prediction result achieved. We choose AUC and LE to show the performance differences while RMSE is omitted as it shows a similar trend as LE. Figure \ref{Figure:paramsensitivity} lays out the results with different parameter settings.

\textbf{Impact of $d$.} The latent dimension $d$ is studied in $\{4,8,16,32,64\}$. As shown in Figure \ref{Figure:paramsensitivity}.(a)-(b), $d$ has an apparent effect on the performance of QFM and QNFM. When $d$ is relatively small (i.e., $d=4$ or $d=8$), it limits the expressiveness of QFM and QNFM, leading to unsatisfactory prediction results on all three datasets. Both models generally benefit from a relatively large value of $d$, but the performance improvement tends to become less significant when $d$ reaches a certain scale, which is $16$ in most of our cases. 

\textbf{Impact of $l$.} We investigate the impact of $l\in \{1,2,3,4,5\}$ in QNFM. Though stacking more deep layers may intuitively help achieve better prediction performance, QNFM barely shows visible improvements in AUC and LE when $l$ increases from $1$ to $3$ with subtle (i.e., 0.0001-level) performance fluctuations. When $l$ reaches $4$ or $5$, the performance of QNFM clearly deteriorates due to overfitting issues caused by excessive deep layers. Hence, setting $l=1$ in QNFM is the most economical option for obtaining advantageous performance with minimal training difficulties.

\textbf{Impact of $\rho$.} In this study, we vary the dropout ratio $\rho$ in $\{0.1, 0.2, 0.3, 0.4, 0.5\}$. Figure \ref{Figure:paramsensitivity}.(d) suggests that the best results are obtained when $\rho$ is between $0.1$ and $0.3$. After increasing the value of $rho$ to $0.4$ and $0.5$, QNFM experiences a significant decrease in performance on Criteo and Avazu. This is because that the huge amount and diversity of instances in these two industry-level datasets can be viewed as an inherent regularization term during training, so a larger $\rho$ will result in underfitting.

\begin{table}[t]
\small
\caption{Ablation test results with different model architectures, where ``$\downarrow$" marks a severe (over $2\%$) performance drop.}
\vspace{-0.3cm}
\centering
\renewcommand{\arraystretch}{1.0}
\setlength\tabcolsep{1.75pt}
  \begin{tabular}{|c|c|c|c c c|}
    \hline
     Model & Metric & Architecture & Trivago & Criteo & Avazu \\
    \hline
   \multirow{6}{*}{QFM} & \multirow{3}{*}{AUC} & Default & \textbf{0.8543} & \textbf{0.8025} & \textbf{0.7739}\\
   & & Dot Product-based FI & { }0.8217$\downarrow$ & 0.7902 & 0.7708 \\
   & & One-Way FI & 0.8476 & 0.7981 & 0.7721 \\
   \cline{2-6}
    & \multirow{3}{*}{LE} & Default & \textbf{0.4696} & \textbf{0.4492} & \textbf{0.3847}\\
   & & Dot Product-based FI & { }0.5154$\downarrow$ & 0.4522 & 0.3860 \\
   & & One-Way FI & 0.4724 & 0.4503 & 0.3852 \\
   \hline
   \hline
   \multirow{8}{*}{QNFM} & \multirow{4}{*}{AUC} & Default & \textbf{0.8838} & \textbf{0.8062} & \textbf{0.7758}\\
   & & Pure Element-wise IP & { }0.8616$\downarrow$ & { }0.7879$\downarrow$ & 0.7667 \\
   & & One-Way IP & 0.8821 & 0.8024 & 0.7743\\
   & & Removing RC & 0.8769 & 0.8010 & 0.7714 \\
   \cline{2-6}
    & \multirow{4}{*}{LE} & Default & \textbf{0.4323} & \textbf{0.4454} & \textbf{0.3823}\\
   & & Pure Element-wise IP & { }0.4419$\downarrow$ & { }0.4579$\downarrow$ & { }0.3905$\downarrow$ \\
   & & One-Way IP & 0.4332 & 0.4489 & 0.3845\\
   & & Removing RC & 0.4356 & 0.4523 & 0.3862 \\
     \hline
    \end{tabular}
\label{table:ablation}
\vspace{-0.7cm}
\end{table}

\subsection{Importance of Quaternion Components (RQ4)}\label{sec:ablation}
To better understand the benefits from the major quaternion components proposed in our models, we conduct ablation test on different degraded versions of QFM and QNFM. Each variant removes or modifies one key component from the model, and the corresponding results on three datasets are reported in Table \ref{table:ablation}. Similar to Section \ref{sec:sensitivity}, AUC and LE are used for demonstration. In what follows, we introduce these variants and analyze their effect respectively.

\textbf{Dot Product-based Feature Interaction (FI).} We replace the inner Hamilton product in QFM with the straightforward dot product between two quaternion cores in the same channel, i.e., Eq.(\ref{eq:product_2}) $\rightarrow \mathbf{r}_i^{\top}\mathbf{r}_j + \mathbf{a}_i^{\top}\mathbf{a}_j\mathbb{I} + \mathbf{b}_i^{\top}\mathbf{b}_j\mathbb{J} + \mathbf{c}_i^{\top}\mathbf{c}_j\mathbb{K}$. On this occasion, QFM actually models pairwise feature interactions in a similar way to the plain FM, where the only difference is that QFM separates the interaction contexts into four channels. Without the infused expressiveness of Hamilton product, significant performance decrease is observed on three datasets, especially on Trivago which has the highest data scarcity.

\textbf{One-Way Feature Interaction (FI).} This is a variant of QFM that simplifies the two-way feature interaction paradigm in Eq.(\ref{eq:twoway_interaction}) to one-way feature interaction, i.e., Eq.(\ref{eq:twoway_interaction}) $\rightarrow \sum_{i=1}^{n} \sum_{j=i+1}^{n} (\mathbf{v}_i^{\diamond} \otimes \mathbf{v}_j^{\diamond})$ where feature $i$ and $j$ only interact with each other once. Consequently, QFM suffers from the obviously inferior performance, which shows that by taking advantage of the non-commutative property of Hamilton product, the two-way FI effectively enriches the learned interaction information for accurate prediction.

\textbf{Pure Element-wise Interaction Pooling (IP).} We use the pure element-wise vector multiplication as in \cite{he2017neural} for QNFM instead of the element-wise Hamilton product to compute $\widetilde{\mathbf{v}}^\diamond$, i.e., Eq.(\ref{eq:product_3}) $\rightarrow \mathbf{r}_i\!\circ \mathbf{r}_j \!+ \mathbf{a}_i\!\circ \mathbf{a}_j\mathbb{I} \!+ \mathbf{b}_i\!\circ \mathbf{b}_j\mathbb{J} \!+ \mathbf{c}_i\!\circ \mathbf{c}_j\mathbb{K}$. Compared with Hamilton product-based pooling that encodes more semantics of feature interactions, the simplistic element-wise IP yields worse performance, and a severe performance drop on LE (over $2\%$) is observed on three datasets.

\textbf{One-Way Interaction Pooling (IP).} In this QNFM variant, by setting Eq.(\ref{eq:pooling}) $\rightarrow \sum_{i=1}^{n} \sum_{j=i+1}^{n} (\mathbf{v}_i^{\diamond} \odot \mathbf{v}_j^{\diamond})$, we downgrade the original asymmetrical quaternion IP operation to one-way IP. Though it can better model feature interactions than pure element-wise IP, there is still a decrease in prediction accuracy. Hence, our proposed asymmetrical quaternion IP is able to make full use of all the contexts from feature interactions, thus ensuring optimal performance in prediction tasks.

\textbf{Removing Residual Connections (RC).} We construct another variant of QNFM by removing the residual connections in the quaternion feed-forward network defined in Eq.(\ref{eq:QFFN}). Apparently, without residual connections, we find that the performance of QNFM gets worse on three datasets. Presumably this is because information in lower layers (i.e., the output generated by the pooling layer) cannot be easily propagated to the final layer, and such information is highly useful for making predictions, especially on our experimental datasets with a large amount of sparse features.

\begin{figure}[t!]
\centering
\begin{tabular}{cc}
	\vspace{-0.1cm}\hspace{-0.3cm}\includegraphics[width=1.9in]{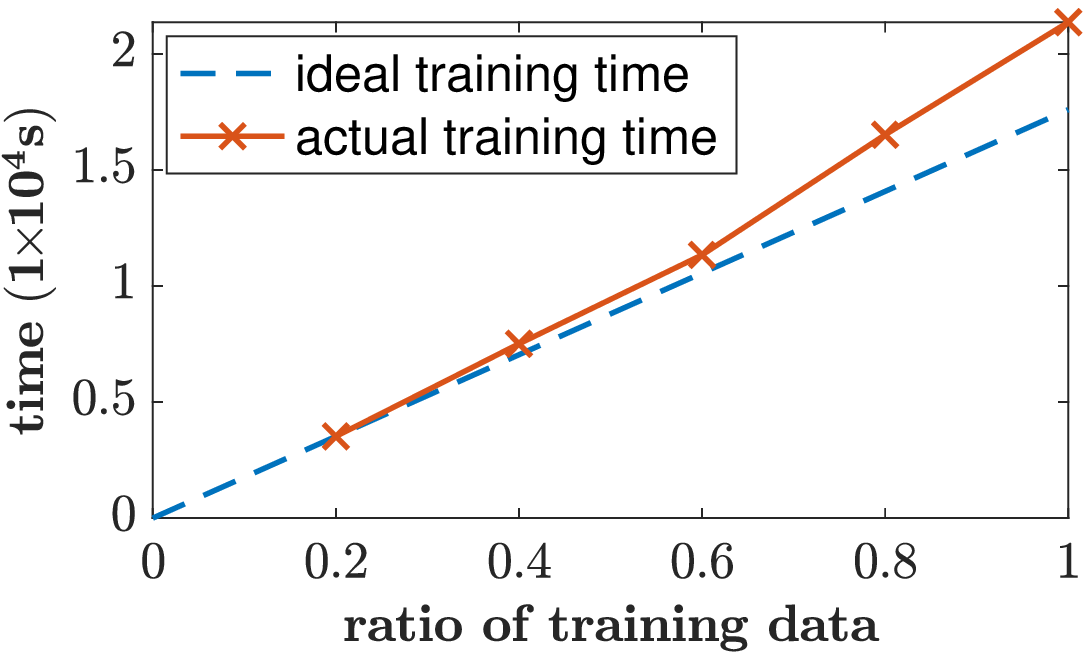}
	&\hspace{-0.8cm}\includegraphics[width=1.9in]{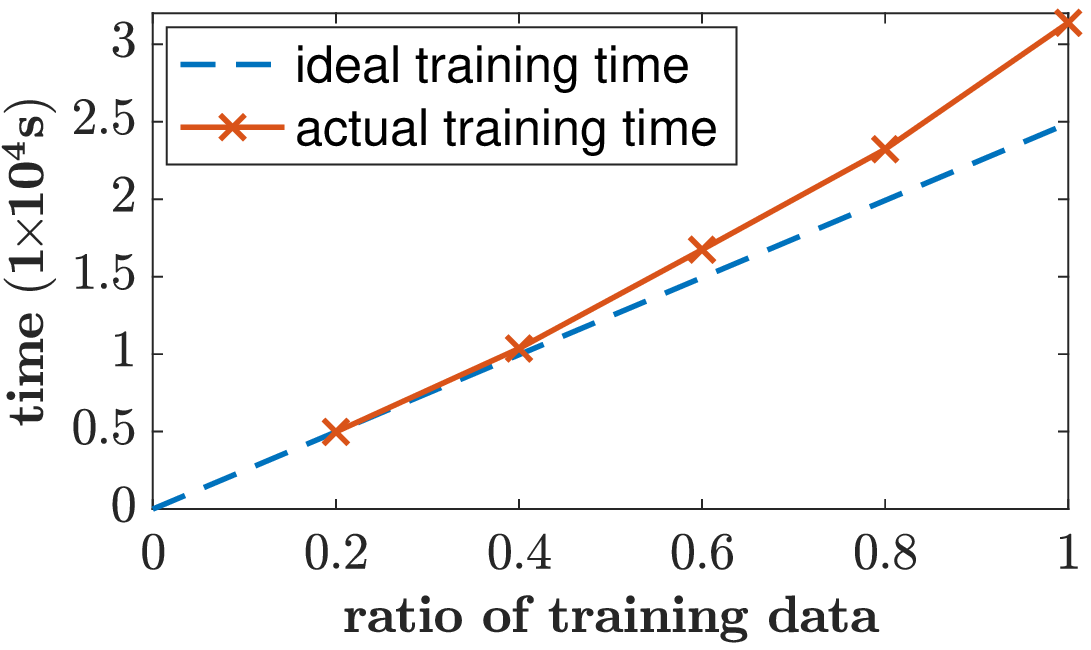}\\
	\footnotesize{(a) QFM} & \hspace{-0.5cm}\footnotesize{(b) QNFM}\\
\end{tabular}
\vspace{-0.3cm}
\caption{Training time of QFM and QNFM w.r.t. varied data proportions.}
\label{Figure:train}
\vspace{-0.5cm}
\end{figure}

\subsection{Training Efficiency and Scalability (RQ5)}
Due to the importance of practicality in real-life applications of predictive models, we validate the training efficiency and scalability of both QFM and QNFM in this section. When all hyperparameters in the network are fixed, the training time for SeqFM is expected to be linearly associated with the number of training samples. We vary the proportions of the training data in $\{0.2, 0.4, 0.6, 0.8, 1.0\}$, and then report the corresponding time cost for the model training. It is worth noting that the Criteo dataset is used for scalability test since it contains the most instances. The training time w.r.t. the data size is shown in Figure \ref{Figure:train}. When the ratio of training data gradually extends from $0.2$ to $1.0$, the training time for QFM increases from $0.352\times10^{4}$ seconds to $2.137\times10^{4}$ seconds. Due to the slightly higher model complexity of QNFM, its training time grows from $0.498\times10^{4}$ seconds to $3.138\times10^{4}$ seconds. The overall trend shows that both models' dependency of training time on the data scale is approximately linear. Hence, we conclude that QFM and QNFM is scalable to even larger datasets. 

\section{Conclusion}\label{sec:conclusion}
In this paper, to mitigate the memory-inefficient nature of existing deep FM-based models, we make an innovative exploration on developing quaternion-valued FM variants by stepping into the hypercomplex space. We propose QFM and QNFM, two novel models that support the modelling of intricate feature interactions with heavily reduced parameter consumption. Experiments on industry-level CTR prediction datasets fully demonstrate that our quaternion FM-based models represent an elegant blend of state-of-the-art prediction performance and lightweight parameterization compared with their real-valued counterparts. 


\end{document}